# Alpha Magnetic Spectrometer (AMS02) experiment on the International Space Station (ISS)

Behcet ALPAT
*(INFN Sezione di Perugia, Via A. Pascoli, 06123, Perugia, Italy)*

**Abstract** The Alpha Magnetic Spectrometer experiment is realized in two phases. A precursor flight (STS-91) with a reduced experimental configuration (AMS01) has successfully flown on space shuttle Discovery in June 1998. The final version (AMS02) will be installed on the International Space Station (ISS) as an independent module in early 2006 for an operational period of three years. The main scientific objectives of AMS02 include the searches for the antimatter and dark matter in cosmic rays. In this work we will discuss the experimental details as well as the improved physics capabilities of AMS02 on ISS.

## 1 Introduction

A possible existence of cosmologically large domains of antimatter or astronomical “anti objects” and the nature of dark matter in the universe are fundamental questions of the modern astroparticle physics and cosmology.

The AMS02 experiment thanks to its large acceptance (~0.65 m$^2$) and its particle identification capability, will study these fundamental aspects with unprecedented sensitivity. This requires the measurement of the physical quantities such as particle momentum, charge and velocity with highest possible degree of confidence.

An unambiguous proof of existence of cosmic antimatter would be observation of antinuclei ($Z\geqslant2$) in cosmic rays. An observation even of a single antihelium or heavier nuclei would demonstrate that primordial antimatter indeed exists and it is not too far from us.[1]

The AMS02 will be able to distinguish a single antihelium nuclei among ~$10^9$ estimated background particles over 3 years.

The project is realized in two phases. In June 1998, a baseline configuration of the experiment has flown on the space shuttle Discovery for 10 days mission on 51.7° orbit at altitudes between 320 and 390 km. From this mission (STS-91) we gathered precious information on detector performance in actual space conditions and on possible background sources. AMS01 has also measured, for the first time, with such an accuracy from space, cosmic ray fluxes in GeV region covering almost the whole Earth surface. The detector layout, performance and the physics results of AMS01 during STS-91 flight (AMS-01) are described in detail elsewhere.[2-8]

In this work we will discuss the experimental configuration as well as the physics capabilities of AMS02 on the International Space Station.

## 2 Details of the AMS02 experiment

The AMS02 is a large acceptance, high precision superconducting magnetic spectrometer designed to measure cosmic ray spectra of individual elements with $Z$<~25 up to TeV region. It can also measure the high energy gamma rays up to few hundreds GeV with very good γ source pointing capability. Fig.1 shows the details of AMS02 experiment.

There are a total of 227,300 electronics channels each providing 16 bits of information with event rates up to 2 kHz[9] corresponding to a total raw data rate of over 1 Gbit/s. The DAQ electronics will reduce the event size, through proper filtering, to the allocated downlink data rate of 2 Mbit/s.

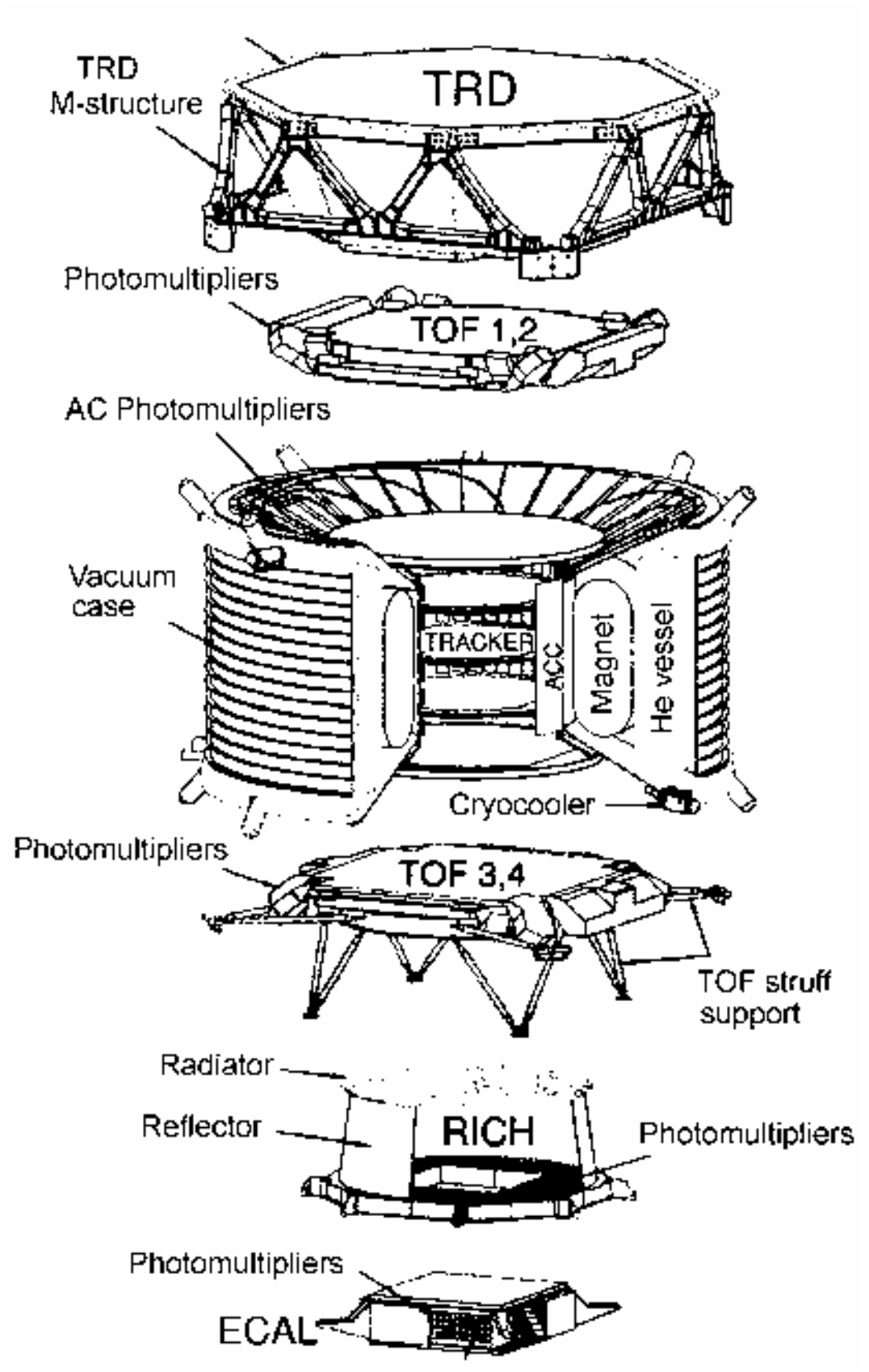

**Fig.1** The exploded view of AMS02.

All electronics and mechanical parts of AMS02 are tested for operation in vacuum. The effect of total ionization dose (up to 6 Gy/year) on all critical components is extensively tested.

The detector has been designed to identify the cosmic rays but at the same time minimizing the multiple scattering and large angle nuclear scattering occurring inside the tracking part of the detector. Large acceptance for antihelium search, good particle rigidity and velocity resolutions as well as their redundant measurements and h/e rejection of $\geqslant 10^6$ were other key parameters for its design.

The AMS02 will weigh 6760 kg and will have a power consumption of 2 kW.

In the following the AMS02 sub-detectors will be described from up to downstream.

### 2.1 Transition Radiation Detector (TRD)

The Transition Radiation Detector is designed to separate e/p signals to distinguish $e^+$ and $\overline{p}$ from relative backgrounds (p and $e^-$ respectively) with a rejection factor of $10^3$~$10^2$ in the energy range from 10~300 GeV.[10] This rejection factor combined with the Electromagnetic Calorimeter will provide an overall $e^+$/p rejection factor of $10^6$ at 90 % of $e^+$ efficiency.

The detector consists of 20 layers of 6 mm diameter straw tubes alternating with 20 mm layers of 10 μm polyethylene/polypropylene fiber radiator. The tubes are filled with a 80% ÷ 20% mixture of Xe ÷ $CO_2$ at 1 bar from a recirculating gas system designed to operate in space for > 3 years. The wall material of straw tubes is a 72 μm kapton foil. The upper and lower 4 layers run in the $x$ direction (parallel to AMS02 magnetic field) while central layers run in the perpendicular $y$ direction to provide bi-dimensional tracking and particle identification.

The TRD performance has been measured at CERN in test beams (p, $e^-$, $\mu^-$, $\pi^-$) with energies in the range 3 to 250 GeV/$c$.[11] In Fig.2 is shown the proton rejection as a function of the particle energy.

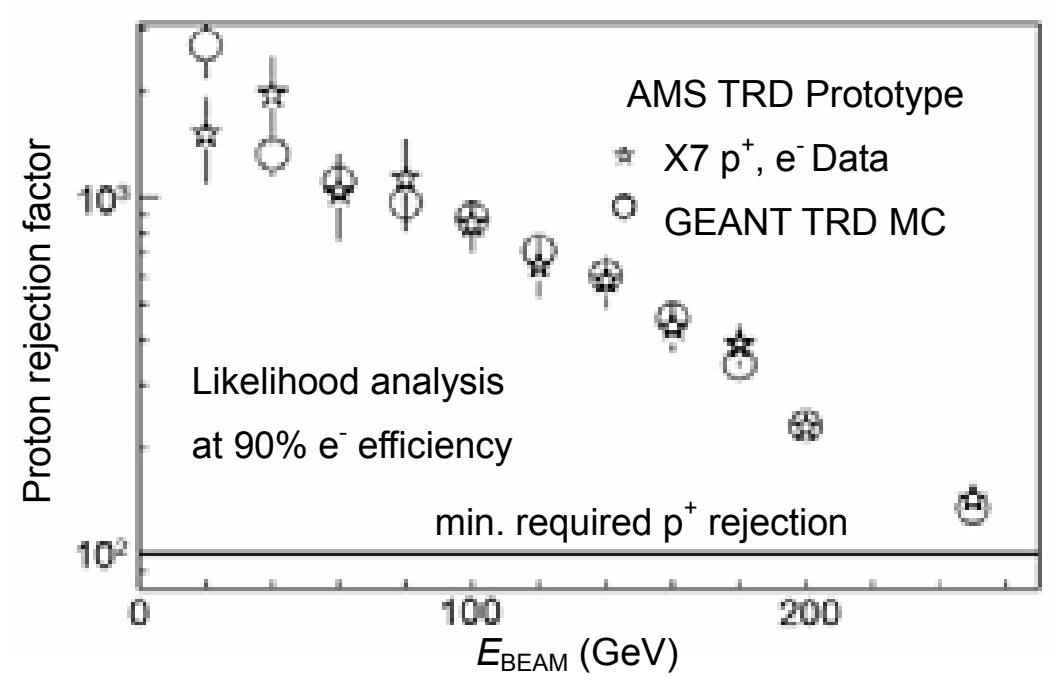

**Fig.2** The proton rejection versus impinging electron energy (see Ref.[11] for details).

### 2.2 Time of Flight (ToF) system

The ToF system is designed to provide fast (first level) trigger to the experiment, measurement of time of flight of the particles traversing the detector with up/down separation better than $10^{-8}$, the measurement of the absolute charge of the particle (in addition to d$E$/d$X$ measured from the Silicon Tracker) and the identification of electrons and positron from antiprotons and protons up to 1~2 GeV. The expected overall time resolution is ≈140 ps for protons and better for heavier cosmic ray nuclei (see Fig.3 for intrinsic resolutions).[12]

It consists of four scintillator planes (see Fig.4) read by a total number of 144 Hamamatsu R5946 phototubes. The design of AMS02 ToF system is determined by the constraint of operation in 0.1~0.3 T stray magnetic field. The choice of PM and the shape of light guides are driven by this issue. Light guides are indeed tilted to reduce the angle between the magnetic field and PM axis.

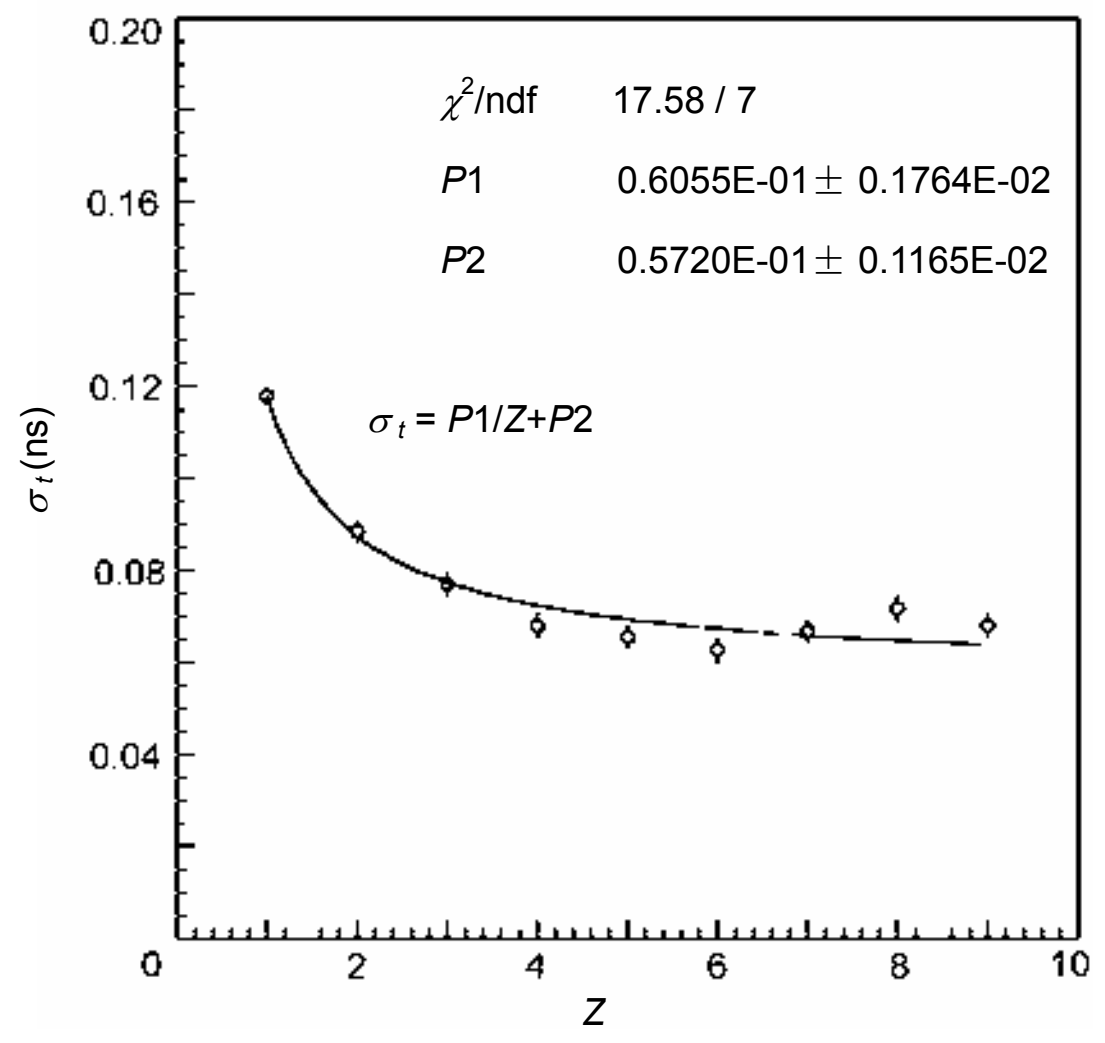

**Fig.3** Average intrinsic time of flight resolution (vertical scale in ns) as a function of particle charge(see Ref. [13] for details).

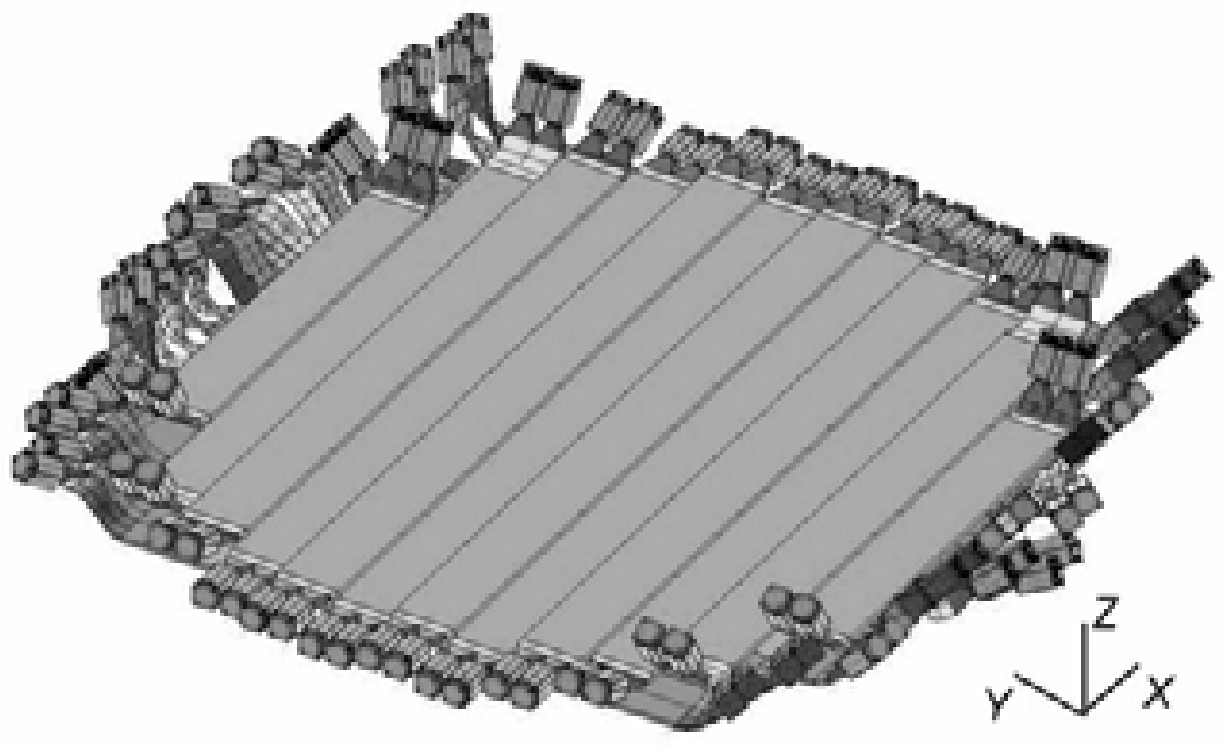

**Fig.4** Scintillator paddles with PM and light guides on a complete ToF counter.

## 2.3 Superconducting Magnet (SCM)

One of the challenging features of the AMS02 detector is its strong superconducting magnet. It is the first large superconducting magnet used in space and it has a bending power of $B{\cdot}L^2 \approx 0.8$ Tm$^2$ [14] which will be essential to perform a sensitive search for antimatter ($\overline{\mathrm{He}}$) in the rigidity ($p/Z$) range from 0.1 GV to several TV.

The magnet consists of 2 dipole coils together with 2 sets of smaller racetrack coils (see Fig.5) with a total cold mass of about 2300 kg. The racetrack coils is designed to increase the overall dipole field, to minimize the stray dipole field outside the magnet (max stray field at a radius of 3 m is 4 mT) in order to avoid an undesirable torque on the ISS caused by the interaction with the Earth magnetic field. All coils are wound from high purity aluminum-stabilized niobium-titanium conductor. The magnet will be operated at a temperature of 1.8 K and cooled by 2500 L of superfluid helium, which should be operational for three years without refilling (optimized for heat losses).[14]

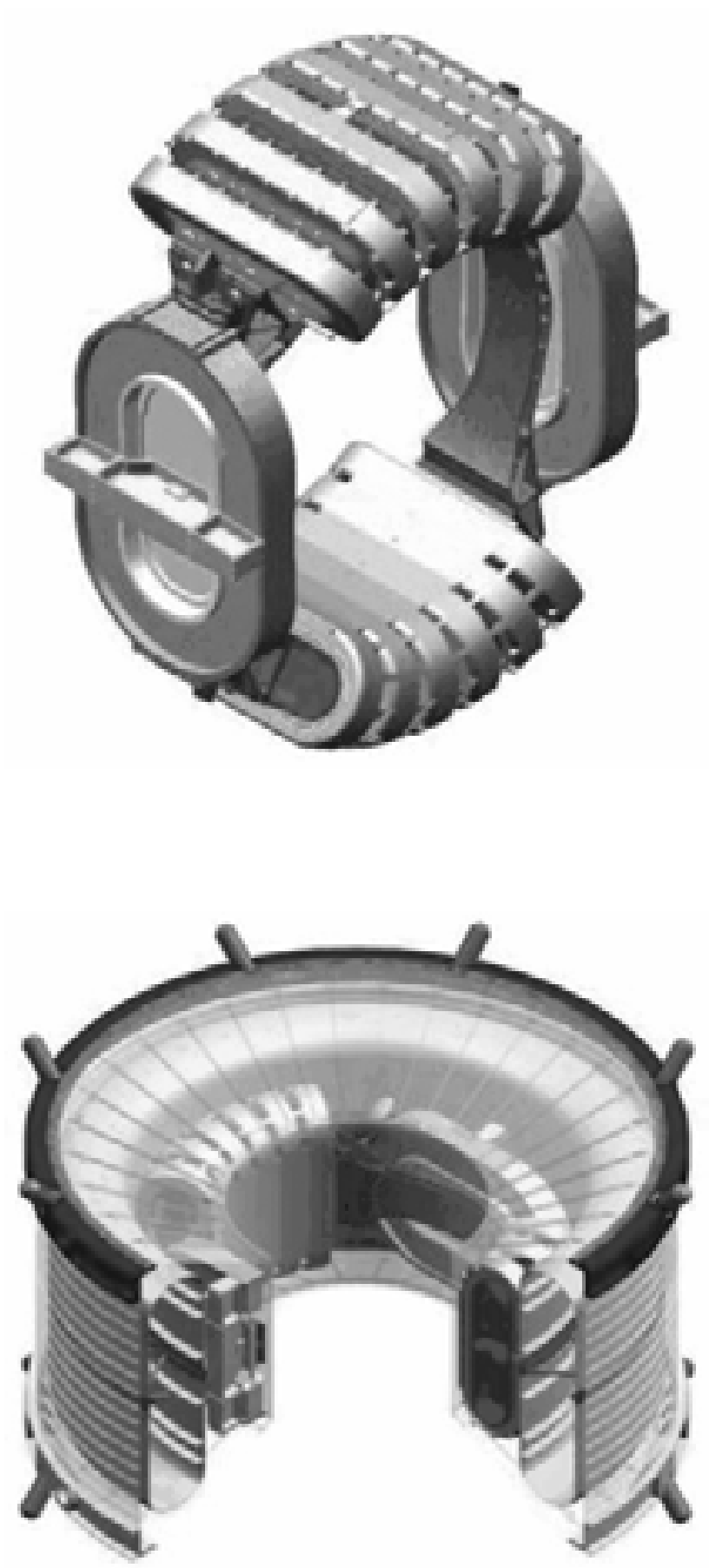

**Fig.5** Upper: the AMS02 superconducting magnet two dipole and 6×2 racetrack coils configuration. Lower: an overall view of the superconducting magnet (see Ref.[14] for details).

## 2.4 Anticoincidence (AC)

The AMS02 anticoincidence system (Fig.6) is designed to assure to trigger only on those particles passing through the aperture of the AMS02 superconducting magnet. It is placed inside the magnet free bore covering the inner surface of the superconducting magnet. AC system consists of thin scintillator slabs readout on both ends by PMs.

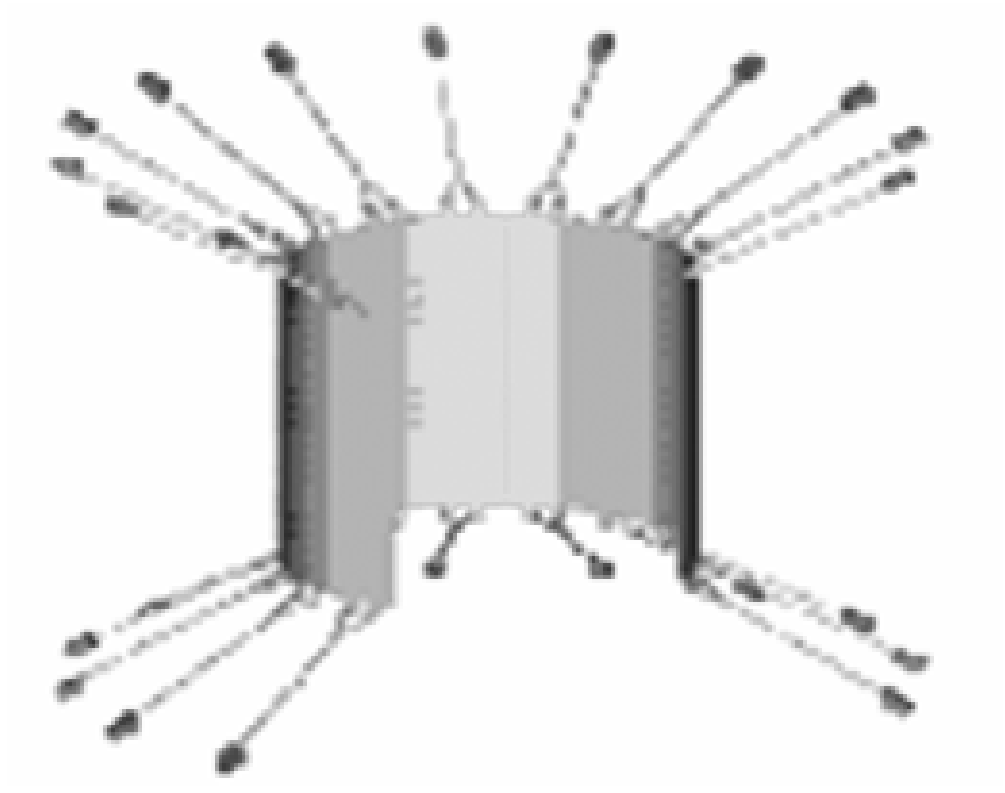

**Fig.6** The anticoincidence system of AMS02.

## 2.5 Silicon Tracker (ST)

The Silicon Tracker of AMS02 is designed to perform high precision measurements of the rigidity, the sign of charge and the absolute charge of the particle traversing it.

The ST consists of 8 thin layers of double-sided silicon microstrip detectors (see Fig.7).There are a total of 192 ladders with variable number of silicon sensors glued together and readout on one extremity by the front-end electronics. The lengths of the ladders vary from 36 cm to about with 60 cm (active part) corresponding to ~6.4 $m^2$ of active double sided surface.[15] Fig.8 shows a detailed design of a ladder and its main components.

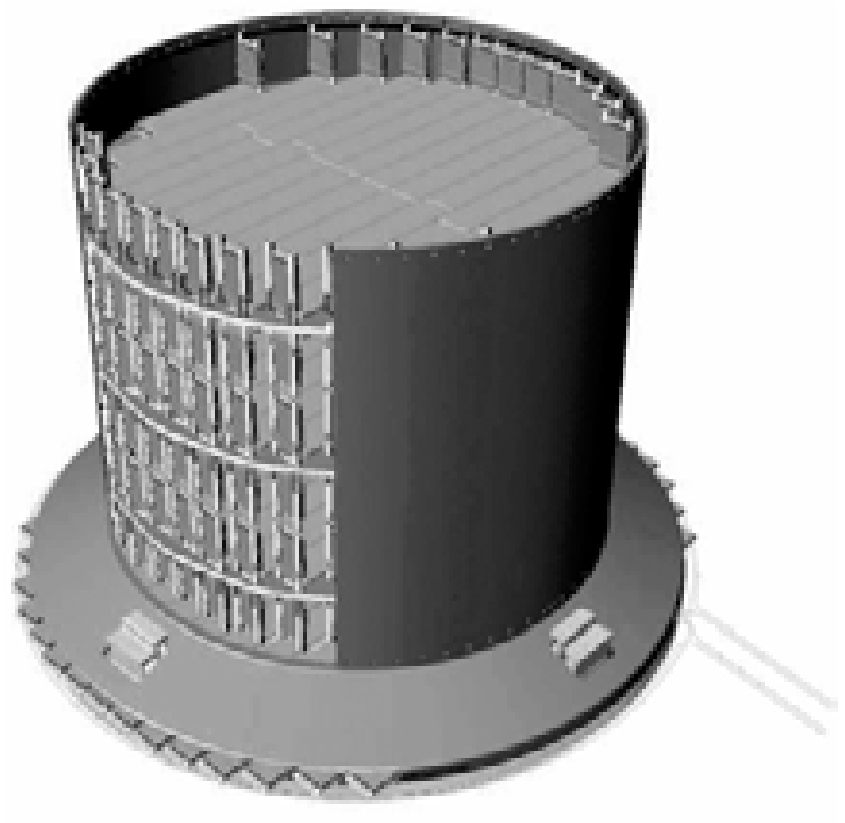

**Fig.7** AMS02 Silicon Tracker 3-D design.

One of the key points in the assembling of long ladders is that it requires high precision in cutting of the sensors and during the ladder assembly. Fig.9 shows the differences between measured (through reference crosses on each sensor) and nominal (perfectly aligned) positions of the sensors of 73 ladders. The r.m.s. of distribution is about 4 μm.

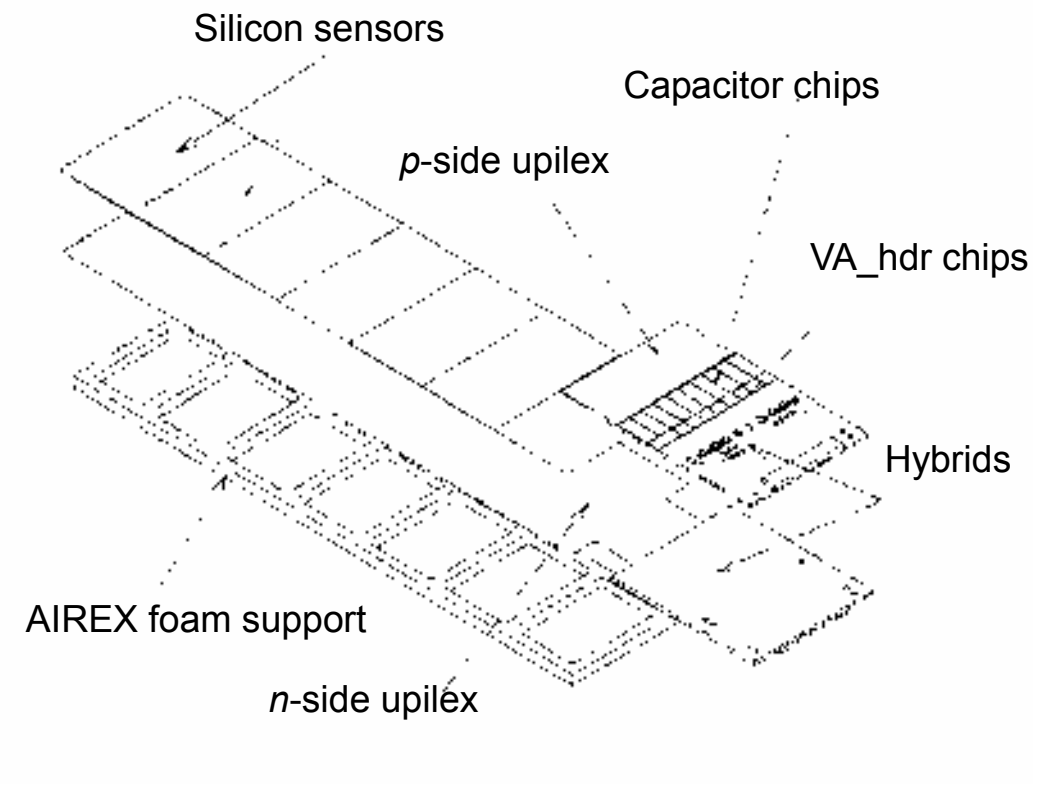

**Fig.8** The AMS02 ladder and its main components.

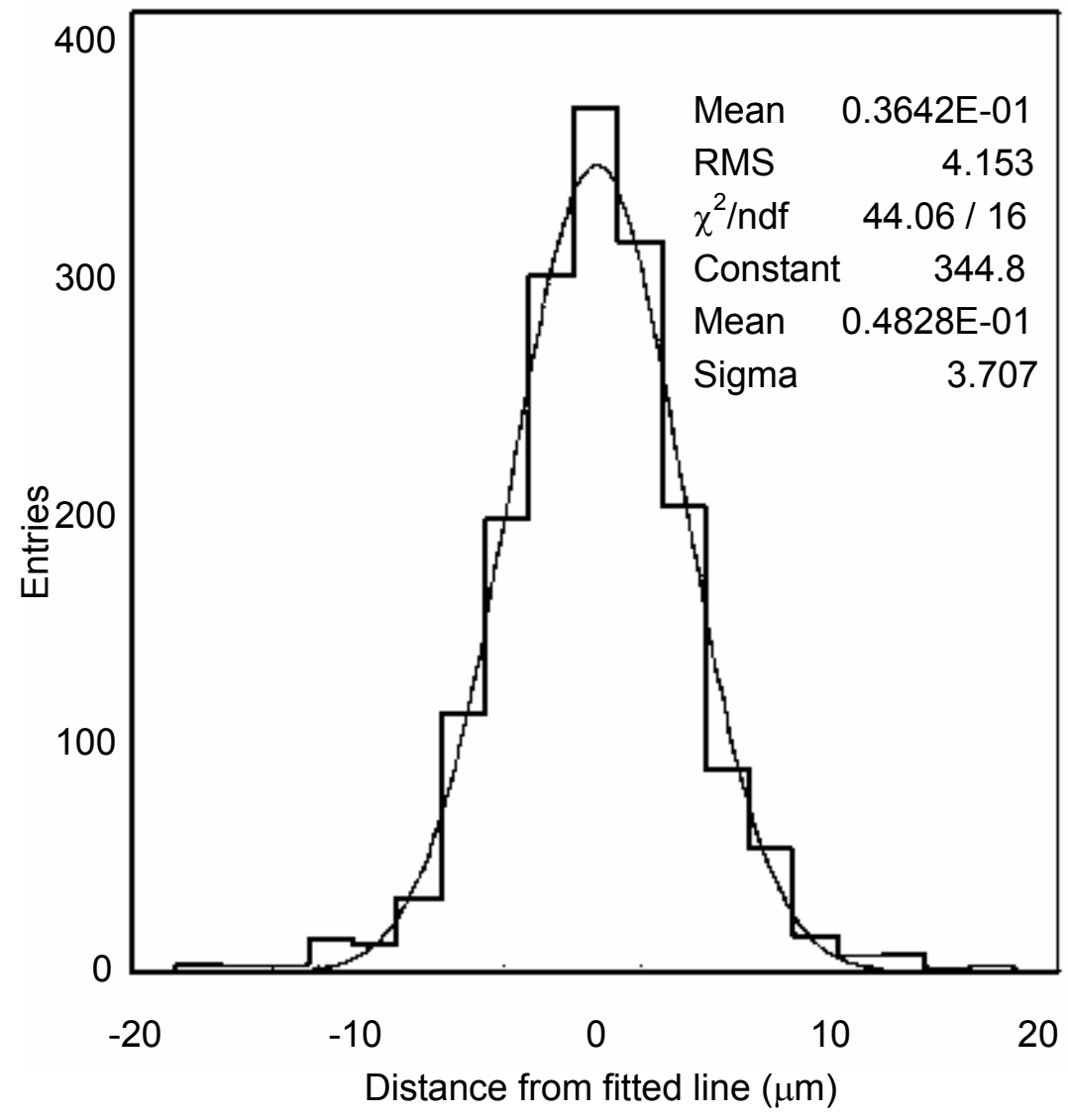

**Fig.9** The residual distribution of 73 ladders measured with 3-D metrology machine. The r.m.s. of the distribution of differences between measured and nominal position of the sensors is ~4μm.

The readout electronics is based on low noise, low power (~0.7 mW/ch), high dynamic range (±70 MIPs) VA_HDR VLSI, preamplifier, shaper, sample and hold circuit[16] connected to the silicon sensors through 700 pF decoupling capacitors.

The performance of the AMS02 ladders has been tested with minimum ionizing particles and with heavy ions.[17] Fig.10 shows the residuals of the reconstructed and expected positions of 400 GeV muons traversing the prototype ladders. The resulting spatial resolutions are 8.5 μm (29.5 μm) and 7.1 μm (22 μm) on bending (non bending) directions for 400 GeV muons and 20 GeV/A helium particles respectively.

Fig.11 shows the d$E$/d$X$ separation capability for the tracker for particles with $Z$ up to 10 (~100 MIPs) using $n$-side information.[17]

In Fig.12 an estimate of the AMS02 proton rigidity resolution (~> 5 hit track) is given.[15]

The planes alignment will be continuously monitored by an IR laser alignment system.

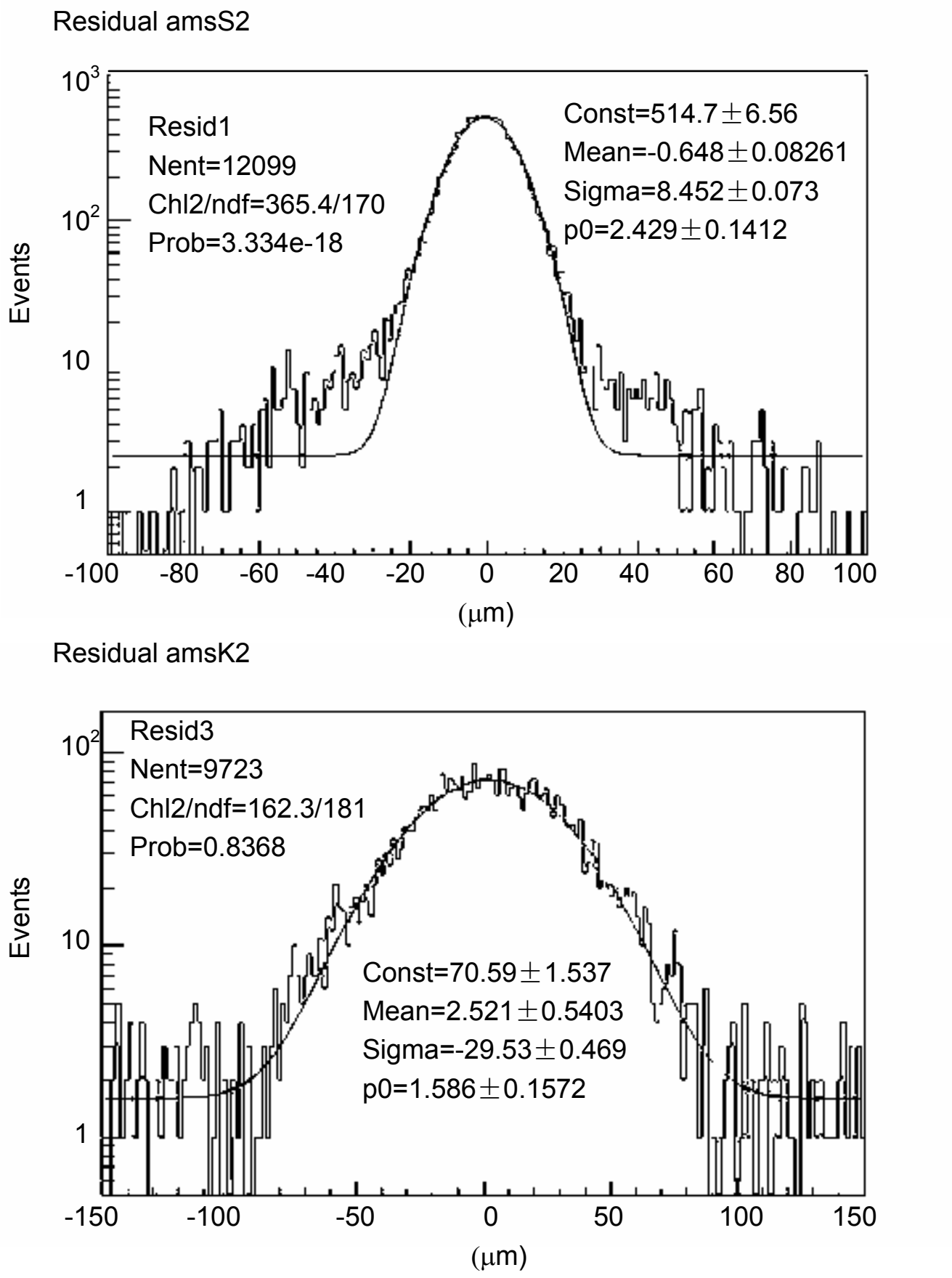

**Fig.10** The distribution of the residuals between measured and expected positions of 400 GeV muons traversing the telescope of the AMS02 prototype ladders. The spatial resolutions are given for p-side (bending plane direction,upper part) and n-side (non-bending,lower part) of ladders.

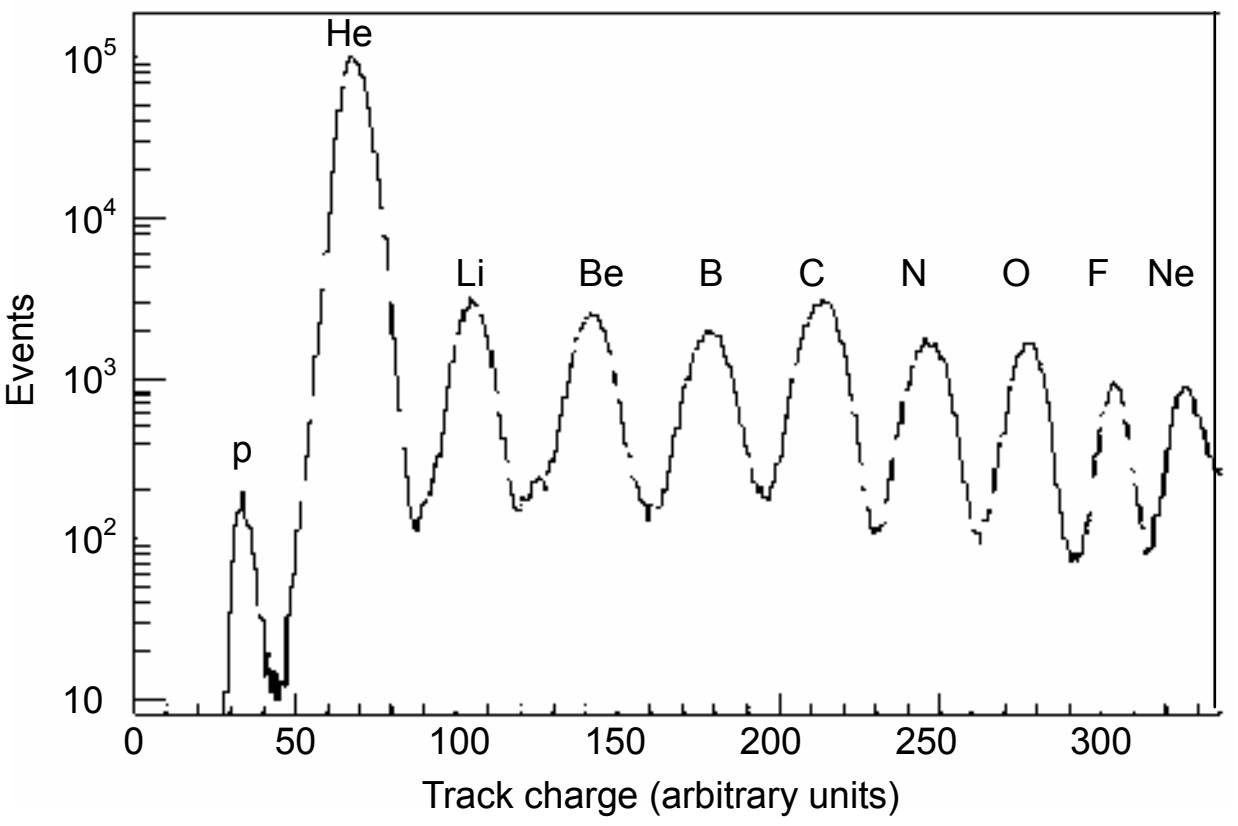

**Fig.11** The $\sqrt{dE/dX}$ distribution on $n$-side obtained in October 2002 beam test at CERN.

The tracker cooling system bases on variable conductive heat-pipes in which the cooling fluid ($CO_2$) runs by the capillary forces. The heat is collected by 3 heat-pipe loops and then driven to the radiators for dissipation (see Fig.13). The system will dissipate 192 W (1 W for each front-end hybrid pair) and fulfill the requirements of operating (survival) temperatures of −10°C to +25°C (−20°C to +40°C) and −10°C to +40°C (−20°C to +60°C) for silicon wafers and front-end hybrids respectively. The requirement for silicon wafer temperature stability per orbit is 3 K and for maximum accepted gradient between any silicon wafer fulfills10 K.[18]

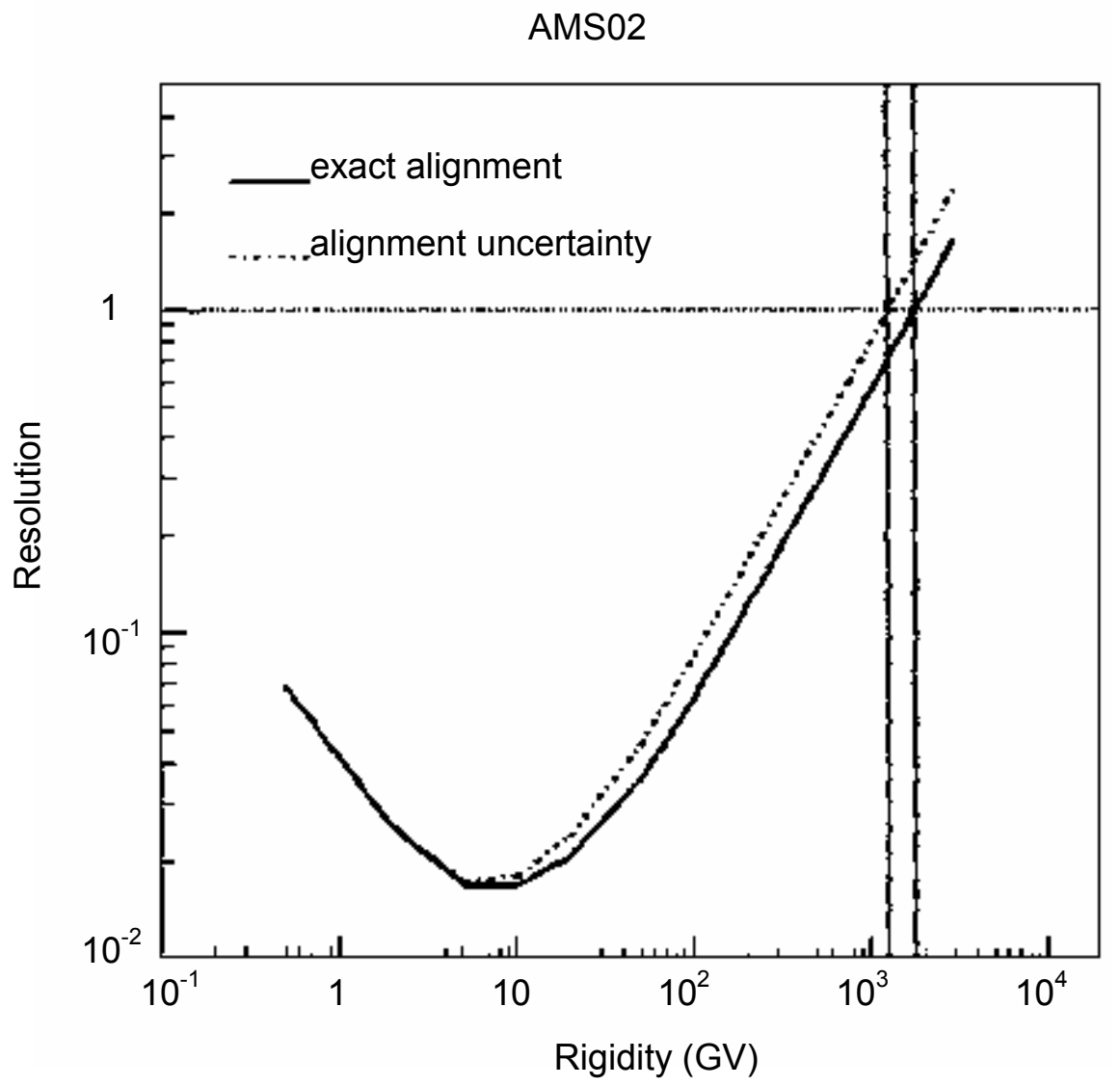

**Fig.12** The simulated rigidity resolution for AMS02 silicon tracker (using five plane tracks). Vertical lines gives the maximum detectible rigidity limits for exact alignment and with alignment uncertainty introduced by displacing the reconstructed track positions by the amount of observed difference between the beam alignment and post-flight (AMS01 on STS-91) metrology (see Ref.[15] for details).

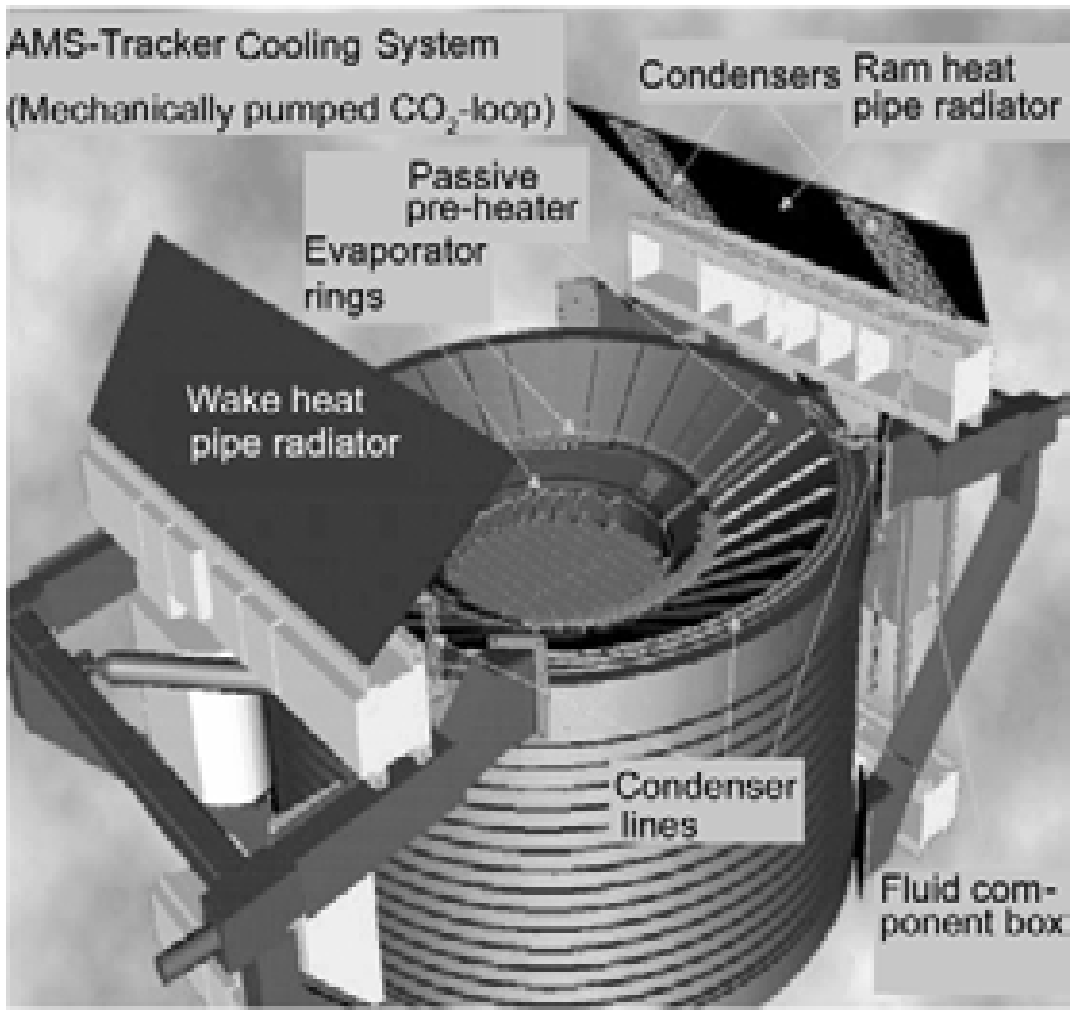

**Fig.13** The AMS02 silicon tracker cooling system details.

The total weight of ST is about 186 kg and the total power consumption is 734 W.

## 2.6 Ring Imaging Cherenkov Detector (RICH)

The particle identification capabilities of the AMS-02 proximity focused RICH detector will improve the confidence in the determination of the sign of the charge, will provide high level of redundancy required for high purity samples of positrons and antiprotons, will perform the identification of isotopes of mass $A$<~15-20, over a momentum range 1 GeV/$c$ $< p/A$ <~12 GeV/$c$ and will identify the chemical composition of elements up to Fe ($Z$~26) to the upper rigidity limit of the spectrometer, $p/Z$ <~1 TV.[19]

The AMS02 RICH detector consists of a plane of radiator material, separated from the detector plane by a drift space in which Cherenkov rings can expand. A detector module includes a matrix of light guides coupled to a PM (Hamamatsu R7900-00-M16) connected to a socket and front-end electronics readout.

The choice made by the Collaboration is to have an Aerogel radiator 3 cm thick with $n$=1.05 in order to cover the momentum interval with a velocity resolution $\Delta\beta/\beta$ of about $1.5\times10^{-3}$. In addition an 0.5 cm thick NaF placed at the central square, corresponding to the hole in the pixels plane, will improve the detection of particles traveling to the central hole direction. Since the Electromagnetic Calorimeter (ECAL) is located just below the RICH photomultipliers plane, the plane is designed with a central hole in order to avoid passive materials in front of ECAL. Fig.14 shows a 3-D design of RICH counter along with the performance test results obtained during the CERN SPS Test beam.[20]

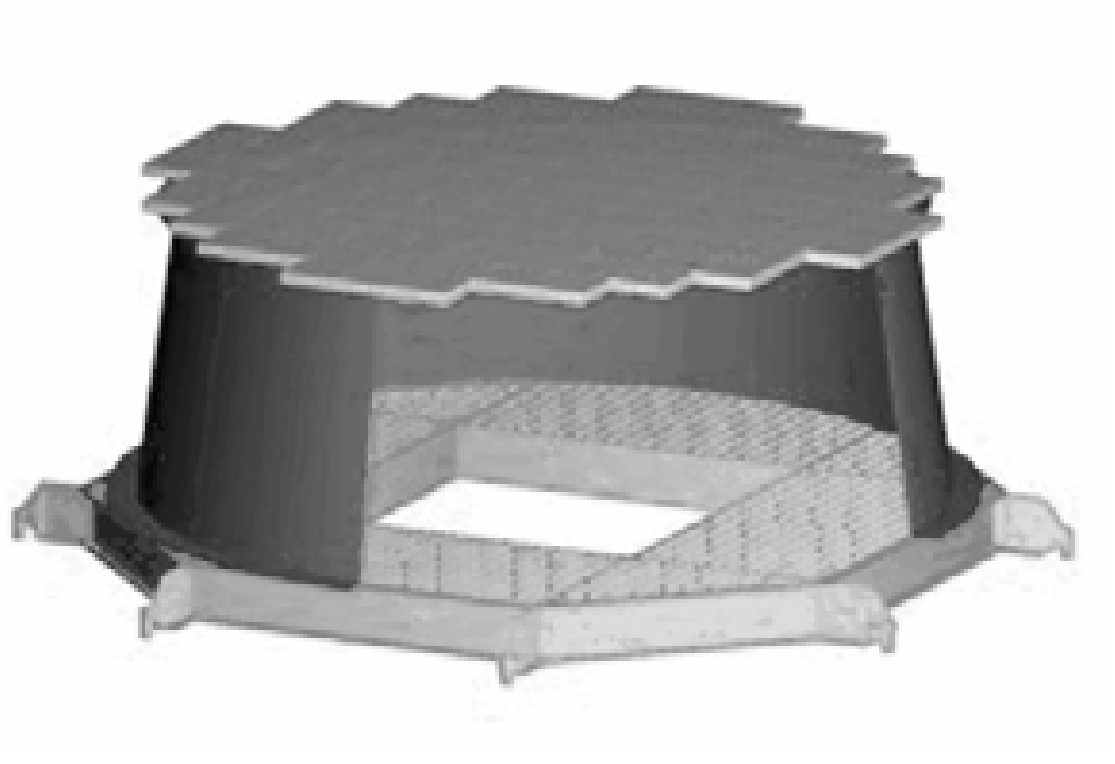

**Fig.14** The 3-D image of the RICH counter with aerogel radiator (top),mirror (cone) and PMs plane(bottom).

Fig.15 is the charge spectrum obtained by the AMS02-RICH prototype under the beam test with 20 GeV/n (per nucleon) ions.

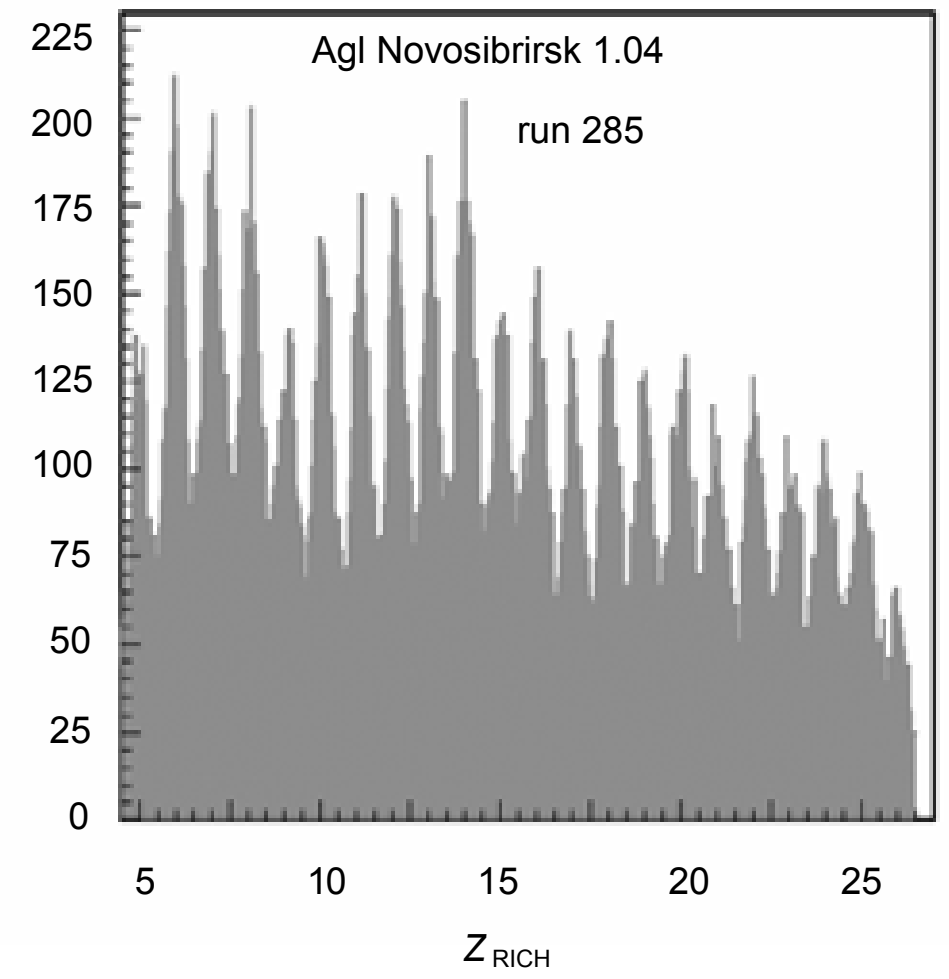

**Fig.15** The charge spectrum obtained from RICH (only) measured with ion beam (20 GeV/$c$ per nucleon) at CERN (see Ref.[20] for details).

## 2.7 Electromagnetic Calorimeter (ECAL)

In order to achieve good e/p separation (design rejection factor of ~$10^5$) which is essential to perform accurate measurement of positron spectra (from few GeV up to ~1 TeV), AMS02 comprises a fine grained sampling electromagnetic calorimeter (ECAL) capable of 3-D imaging of the shower development and of discrimination between hadronic and electromagnetic cascades.[21]

ECAL is a sampling device with a lead-scintillating fibers structure. It has a square parallelepiped shape with 65.8 cm side and 16.5 cm depth. It is segmented in 9 superlayers along its depth and each superlayer, of 18.5 mm total thickness, contains 11 grooved lead foils interleaved with 1 mm diameter scintillating fibers glued with an epoxy resin (average superlayer density of ~6.8 ± 0.3 g/cm$^3$). The calorimeter has a radiation length of about 10mm, total thickness of almost 16 radiation lengths and it allows for 18 samplings in depth (10 in $Y$ and 8 in $X$ views). Fig.16 shows the design of a superlayer.

The performance of this calorimeter has been tested with a full scale prototype under electron and proton beams (3÷100 GeV) at CERN. The energy resolution obtained from these tests is parameterized as $\sigma(E)/E = (11.9 \pm 0.4)\%/\sqrt{E(\mathrm{GeV})} + (2.8 \pm 0.1)\%$ (see Fig.17).

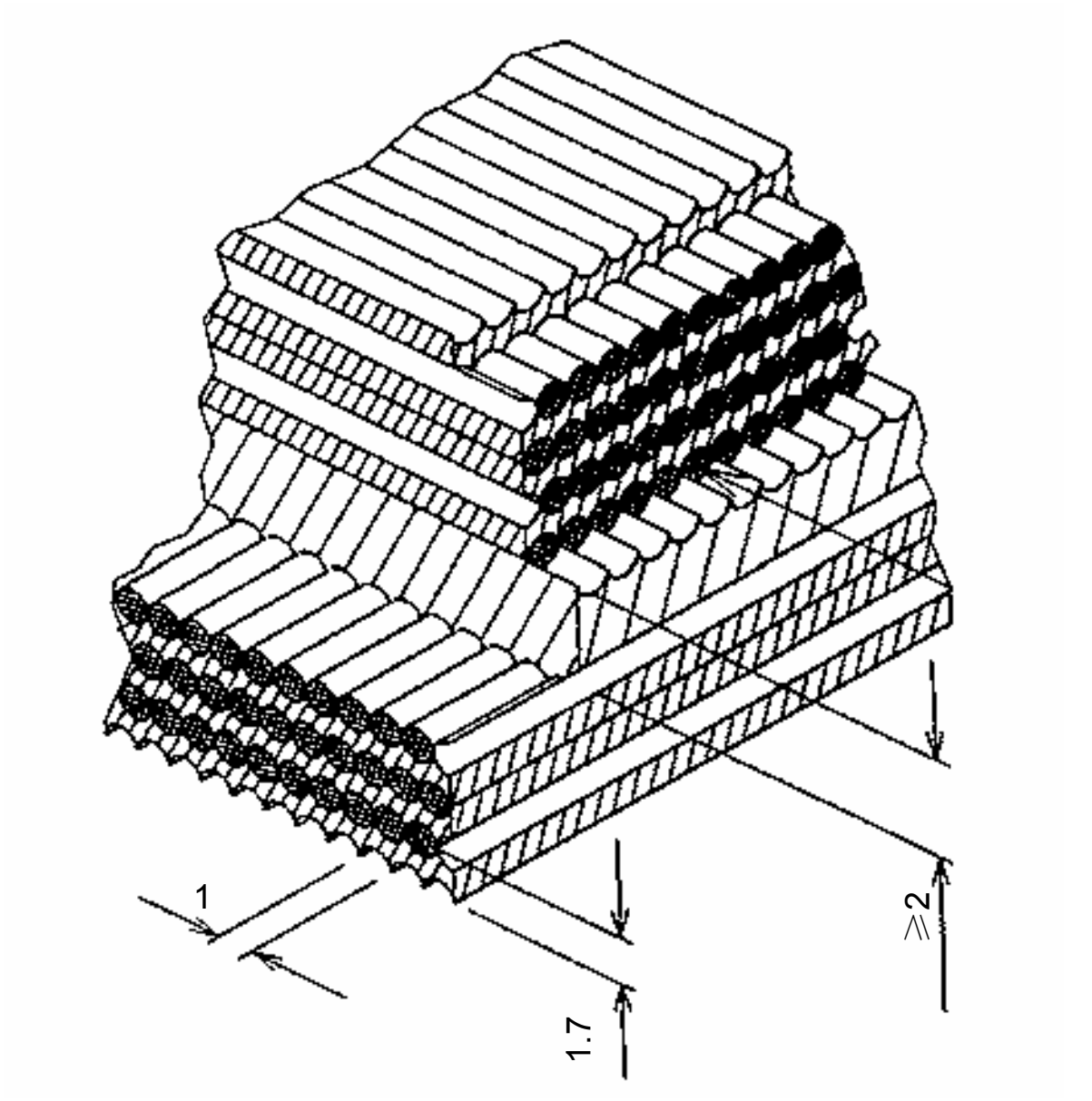

**Fig.16** One superlayer of ECAL with bi-directional placement of scintillating fibers on grooved lead foils.

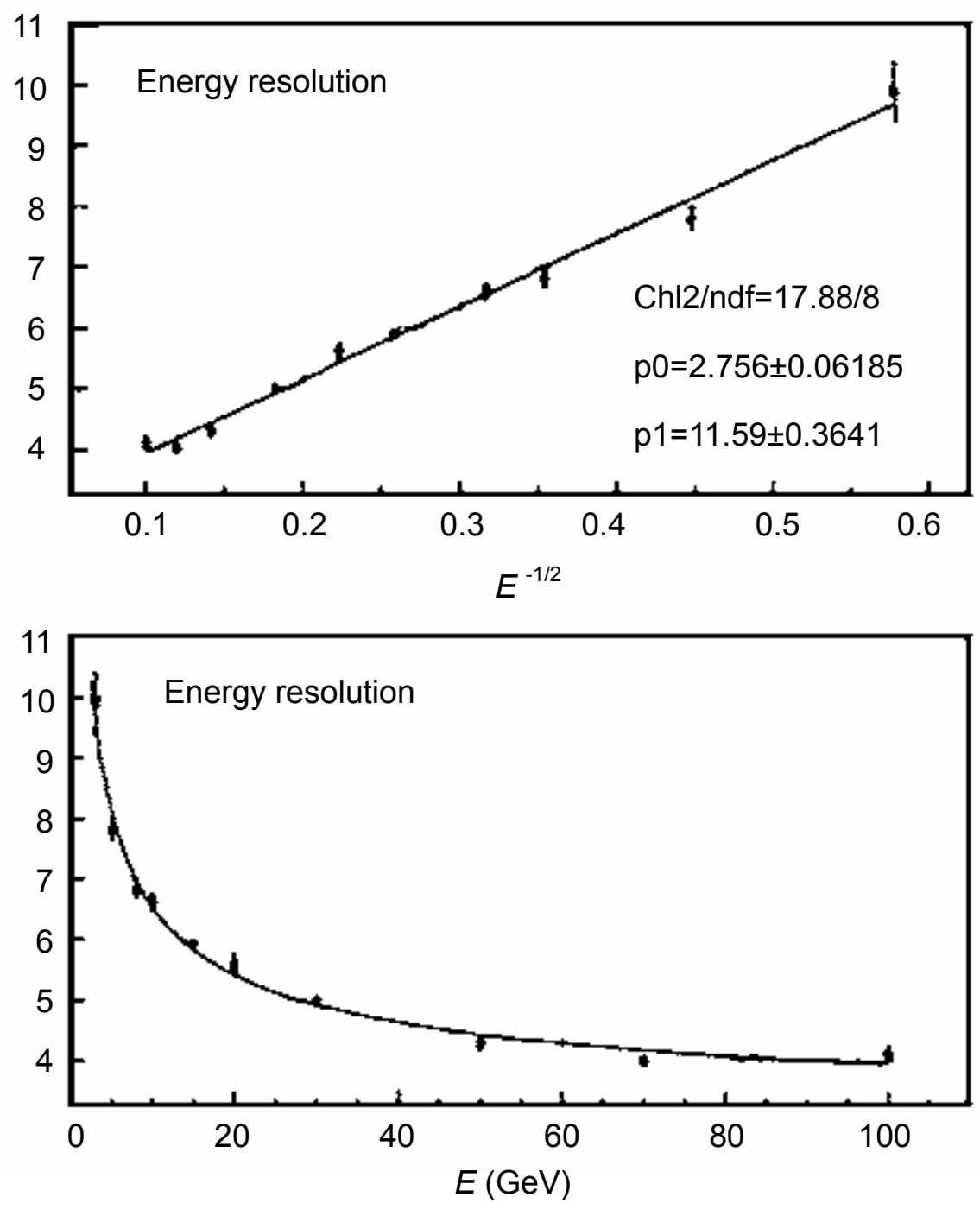

**Fig.17** Energy resolution ( $\sigma(E)/E$ ) is plotted as a function of $E^{-1/2}$ where $E$ is the nominal beam energy (in GeV) (see Ref.[21] for details).

### 2.8 Star Tracker (AMICA)

The main purpose of Astro Mapper for Instrument Check of Attitude (AMICA) is to provide accurate pointing direction for AMS02. AMICA will give a precise measurement of the AMS02 observing direction with a few arc-sec accuracy. The hardware consists of an optics system (f1/2 lens with 75 mm focal length and with a 6.5°×4.8° FoV), an intensified frame-transfer CCD (385×288 pixels) and a baffle to limit the reflections during the daylight part of the orbit. The electronics unit is based on a VME bus which contains the processor (DSP21020) and housekeeping boards.

## 3 AMS02 Physics

### 3.1 Antimatter search

Our region of the universe is certainly dominated by matter. Tiny amount of antiprotons and positrons present in cosmic rays can be explained as secondaries of ordinary matter (protons, electrons and nuclei consisting in protons and neutrons) and gamma rays interacting with interstellar material. The ratio of the excess of baryonic matter over antimatter, is usually given[22] as $\beta=(N_B-N_{\bar{B}})/N_\gamma$ $\approx 6.10^{-10}$ where $N_{B,\bar{B},\gamma}$ are respectively the cosmic number densities of baryons, antibaryons and photons in microwave background radiation (CMBR). At present day, from the direct observation we have $N_\gamma = 411.4/\mathrm{cm}^3$ [23] and $N_B \rangle\rangle N_{\bar{B}}$ (at least in our neighborhood).

According to Sakharov,[24] to generate the baryon asymmetry three principles of baryogenesis should be fulfilled: non conservation of baryonic charge, breaking of C and CP invariance (symmetry breaking between particles and antiparticles) and the deviation from thermal equilibrium. There are several scenarios of baryogenesis which are based on the assumption of C and CP violation and give the cosmological baryon asymmetry with $\beta$=constant and no cosmological antimatter. However if charge symmetry is broken spontaneously, then in different CP domains the universe could be either baryonic or anti-baryonic. The size of these domains may be cosmologically large if after their formation the universe has passed through a period of inflation (exponential expansion).[25] In principle, subsequent baryogenesis can lead to separate regions containing matter and antimatter galaxies.

Recent reviews of (anti)-baryogenesis scenarios can be found in references.[22,26]

It has been shown[27] that the ratio of extragalactic/galactic cosmic rays should increase with energy owing to the fact that the escape rate of the galactic cosmic rays from the galaxy increases with energy. It is also argued that the galactic wind impedes the entrance of extragalactic cosmic rays and the cosmic rays may not propagate towards the Earth from tens of Mpc as required by some models. On the other hand it would not be accurate to estimate from how far the extragalactic nuclei can reach the Earth since we have very limited knowledge about the extragalactic field strength.[28] It is however true stating that more sensitive test for extragalactic antimatter can be done at high rigidities (>hundreds of GV). Fig.18 shows expected AMS02 sensitivity to antihelium nuclei for 3 years of data taking period on ISS.

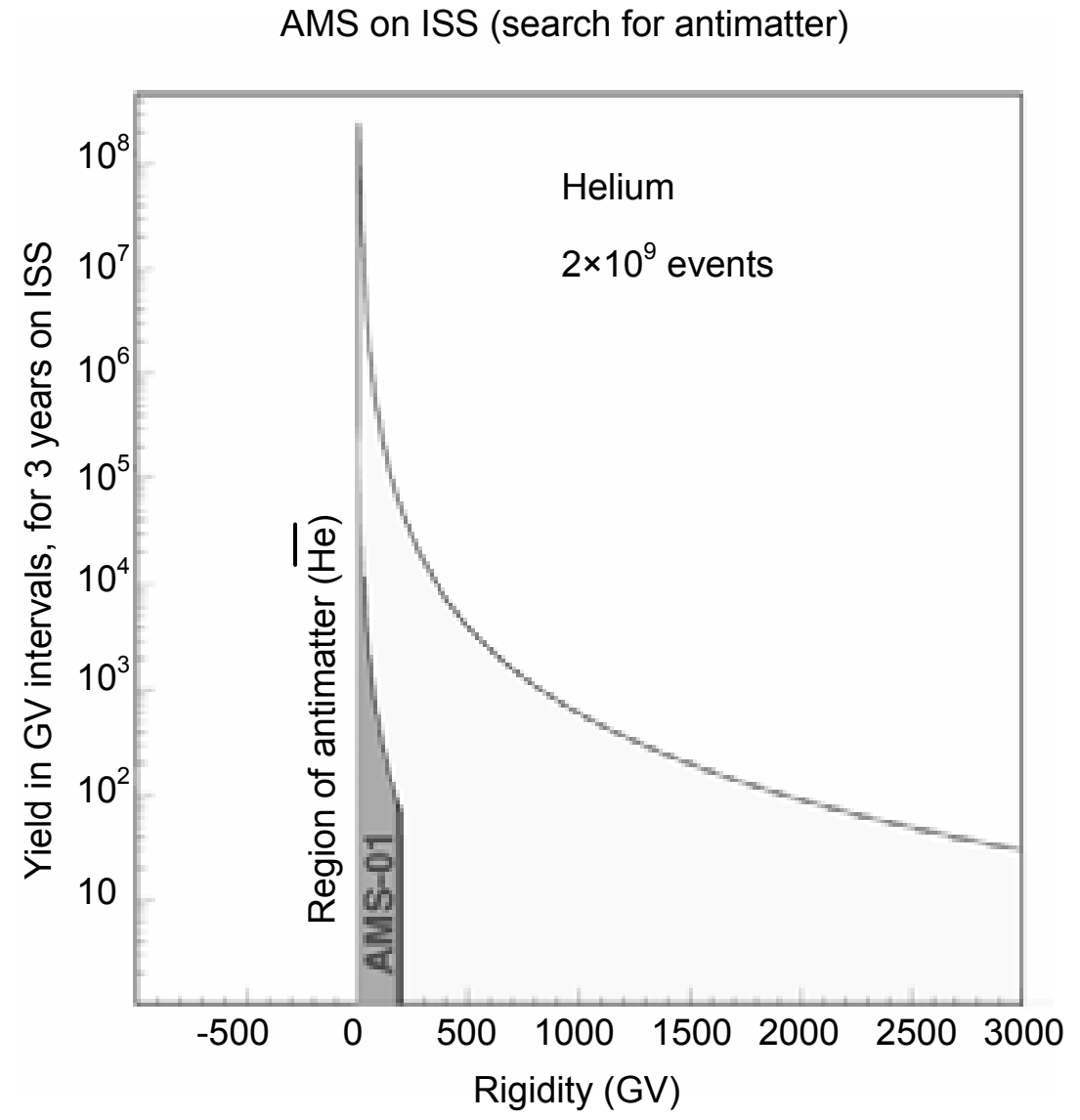

**Fig.18** The simulated sensitivity for antihelium search by AMS02 after 3 years on ISS. The region studied during AMS01 is shown for comparison.

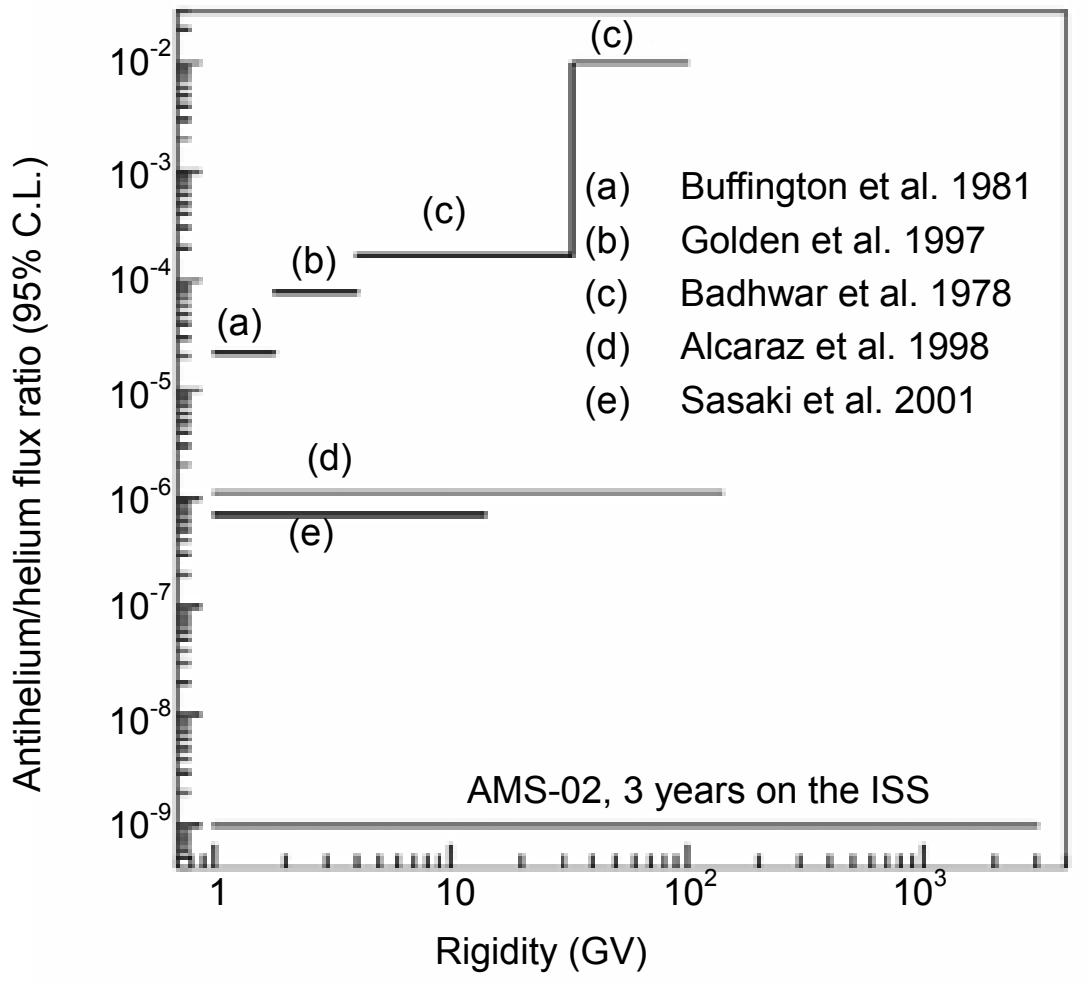

**Fig.19** The expected antihelium/helium ratio as a function of rigidity for AMS02 as well as the experimental data available to day. For AMS01 experimental and AMS02 expected limit the same spectrum for He and $\overline{\text{He}}$ were considered.

Until recently, the search for antinuclei and antiproton has been carried out by stratospheric balloons, and on spacecrafts and no antinuclei was observed.[29-31] The limits on antihelium/helium ratio published by various experiments as well as the simulated sensitivity for AMS02 are shown in Fig.19.

### 3.2 Dark matter

The observation of the rotational velocities of stars in spiral galaxies enables us to calculate the mean density of the matter as a function of the distance from the galactic center. From the recent WMAP measurements of Cosmic Microwave Background (CMB) anisotropies,[32] the total amount of matter is close to the critical density for a flat Universe with $\Omega_{\text{matter}} = 0.27 \pm 0.04$. The contribution of the luminous matter (stars, emitting clouds of gases) is $\Omega_{\text{lum}} < 0.01$ and a precise determination of primeval abundance of deuterium provide strong limits on the value of baryon density $\Omega_{\text{B}} = 0.045 \pm 0.005$.[33] The conclusion from these measurements is that most of the matter is non-luminous and non-baryonic.

The weakly interacting massive particles (WIMPs), postulated in minimal supersymmetric standard model (MSSM) and in other R-parity conserving supersymmetric models, are particularly attractive to explain dark matter's nature. In this framework the lightest supersymmetric particle (LSP), stable neutralino, $\chi$, a neutral scalar boson being also its own antiparticle, is the most quoted candidate.[34]

Indirect signals may be produced by annihilation of neutralinos inside celestial bodies (Earth and Sun) where $\chi$'s have been captured and accumulated. When $\chi\chi$ annihilations take place in the galactic halo specific signals would emerge.[35]

The stable cosmic ray species generated in neutralino annihilations include gamma rays, neutrinos, positrons, antiprotons and antideuterons and, in the same amounts, their counterparts with opposite lepton and baryon numbers.

In the following we will discuss AMS02's capabilities, being the unique experiment[36] to measure all neutralino annihilation products (except neutrinos) with the same apparatus.

#### 3.2.1 Gamma rays

The gamma rays are important tracers to probe the high energy processes in Universe. They travel over the entire Universe along straight lines without significant absorption and transport information about high/extreme energy interactions, objects or events from distant domains. Among proposed dark matter gamma ray sources, there are the Galactic center, the whole Milky Way halo, external galaxies and cosmological sources.

About 20~30% of the energy released in WIMP annihilations goes into gamma rays. Most of them (~90%) are generated in the decay of neutral pions in fragmentation processes.[37,38]

The WIMPs in the galactic halos move with $\beta \sim 10^{-3}$ hence the photons from neutralino annihilation ($\chi\chi \to \gamma\gamma$ or $\chi\chi \to Z\gamma$) are nearly monochromatic with energy of $O(m_\chi)$.[39] Since there is no other known source with a similar behavior a line shape gamma ray signal over the background would be a clear confirmation of the existence of WIMP based dark matter.

A compilation of estimates of flux sensitivities is given in Fig.20 for space borne (upper left part in the Figure) and Earth based (lower right) experiments showing which will be the progress in next decade and what signal levels will be.

Most of the high energy gamma ray data come from EGRET experiment on Gamma Ray Observatory (GRO). After the completion of GRO program (in June 2000) there is no experiment measuring high energy gamma rays in space. In upcoming decade there will be three experiments, AGILE (in 2004[41]), AMS02 on ISS and GLAST (2007[42]) which will be able to cover the energy spectrum region from 20 MeV up to about 300 GeV.

Despite their low flux, high energy gamma rays bring valuable information complementary to those gathered from charged particles. Since they do not deflect on their path during the travel in the (inter)galactic magnetic field, it is possible to point to the galactic or extra galactic source directions. A good source pointing and capability to perform accurate spectral studies require good angular and energy resolutions respectively.

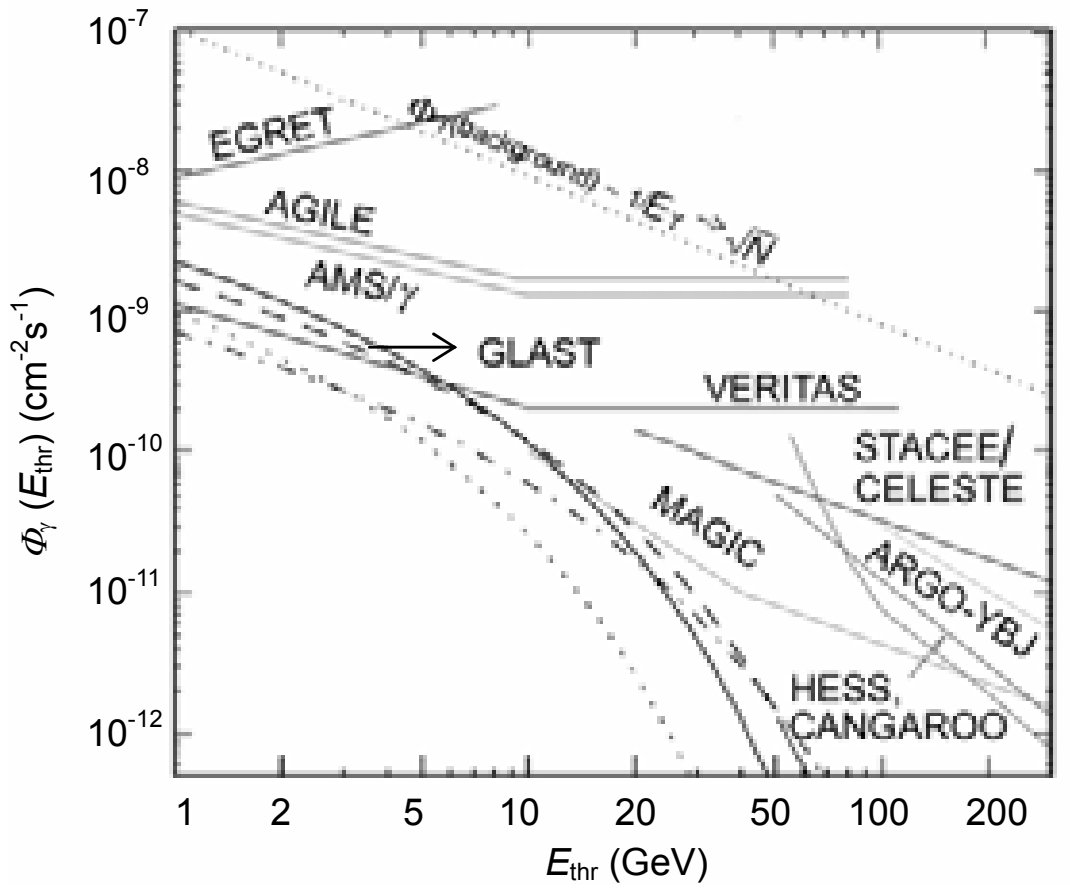

**Fig.20** Estimates of flux sensitivities to four Supersymmetry models for upcoming experiments. Integral photon fluxes $\Phi_{\square}(E_{\mathrm{thr}})$ as a function of threshold energy ($E_{\mathrm{thr}}$) are given for $A0=0$, $\mu > 0$, $mt = 174$ GeV and halo parameter $J_{\mathrm{bar}} = 500$ (see [40] for details).

The AMS02 detector geometry and performance were simulated by using two complementary methods (see Table 1). The conversion mode consists in reconstruction of two tracks ($e^+$, $e^-$) in the tracker created by the interaction of incoming photon with the material present upstream to the first tracker layer (TRD, ToF and support material a total of ~0.23 $X_0$). The detector acceptance reaches a maximum of ~0.058 $m^2 \cdot sr$ in the energy range from 7 to 200 GeV.

Instead, the single photon mode was based on the photon detection in electromagnetic calorimeter only. The signature for this method was the presence of an electromagnetic shower in ECAL and no signal in other sub-detectors. The detector acceptance in this case varies between a maximum of a ~0.065 $m^2 \cdot sr$ around 3 GeV and ~0.040 $m^2 \cdot sr$ at about 1 TeV.

**Table 1** The energy and angular resolutions for high energy gamma ray detection for AMS02 are parameterized for tracker (conversion mode) and electromagnetic calorimeter (single photon mode).[43]

| γ resolutions | Conversion mode (Tracker only) | Single photon mode (ECAL) |
|---|---|---|
| Energy | $\sigma(E)/E = 0.03 \oplus 0.5E(\mathrm{TeV})$ | $\sigma(E)/E = 0.03 \oplus 0.13/\sqrt{E(\mathrm{GeV})}$ |
| Angular | $\sigma_{68}(E) = 0.018^\circ \oplus 0.85^\circ/E(\mathrm{GeV})$ | $\sigma_{68}(E) = 0.9^\circ \oplus 0.85^\circ/E(\mathrm{GeV})$ |

### 3.2.2 Antiprotons

Calculation of secondary antiprotons, due to interactions of cosmic rays with Interstellar material, have greatly improved in recent years[44,45] and have shown that at low energy (below kinematic limit; $T_{\bar{p}} \leqslant 1$ GeV) the secondary spectrum is much flatter than previously believed and fits remarkably well the experimental data.[46] This makes the extraction of supersymmetric signal from the background more difficult. Although the measured antiproton flux gives rather stringent limits on MSSM models with the highest annihilation rates, the experimental upper limits may be used to bound from below the lifetime of hypothetical R-parity violating decaying neutralinos.[47] In some other scenarios with a clumpy halo (which enhances the annihilation rate) there may be the possibility to detect heavy neutralinos through spectral features above several GeV.[48]

Fig.21 shows antiproton/proton ratio given by balloon experiments together with predictions by theoretical calculations. Solid curves are upper and lower limit assuming secondary production only.[49] Dashed curve is a similar calculation by L. Bergstrom and P. Ullio.[50]

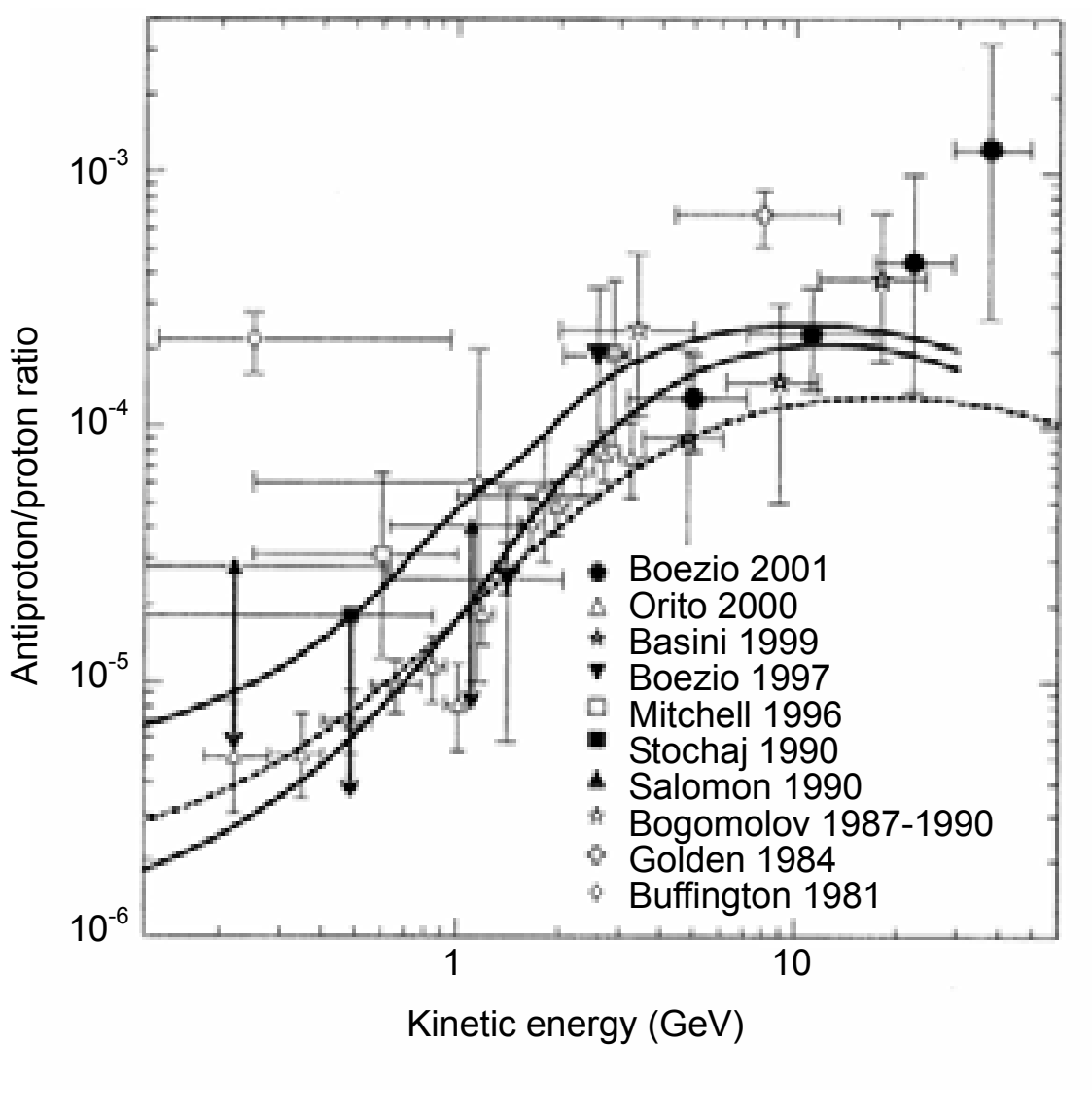

**Fig.21** The antiproton/proton ratio for all experimental data. The solid curves are upper and lower limits for pure secondary production during CR propagation in the galaxy. Dashed curve gives a similar calculation by L. Bergstrom and P. Ullio.[49,50]

Instead, Fig.22 includes few thousands of antiprotons measured during last four decades by balloon borne experiments as well as AMS01. The errors (both statistic and systematic) are larger at higher energies. The figure shows also the AMS02 capability to extend the energy range up to about 300 GeV with much higher statistics.

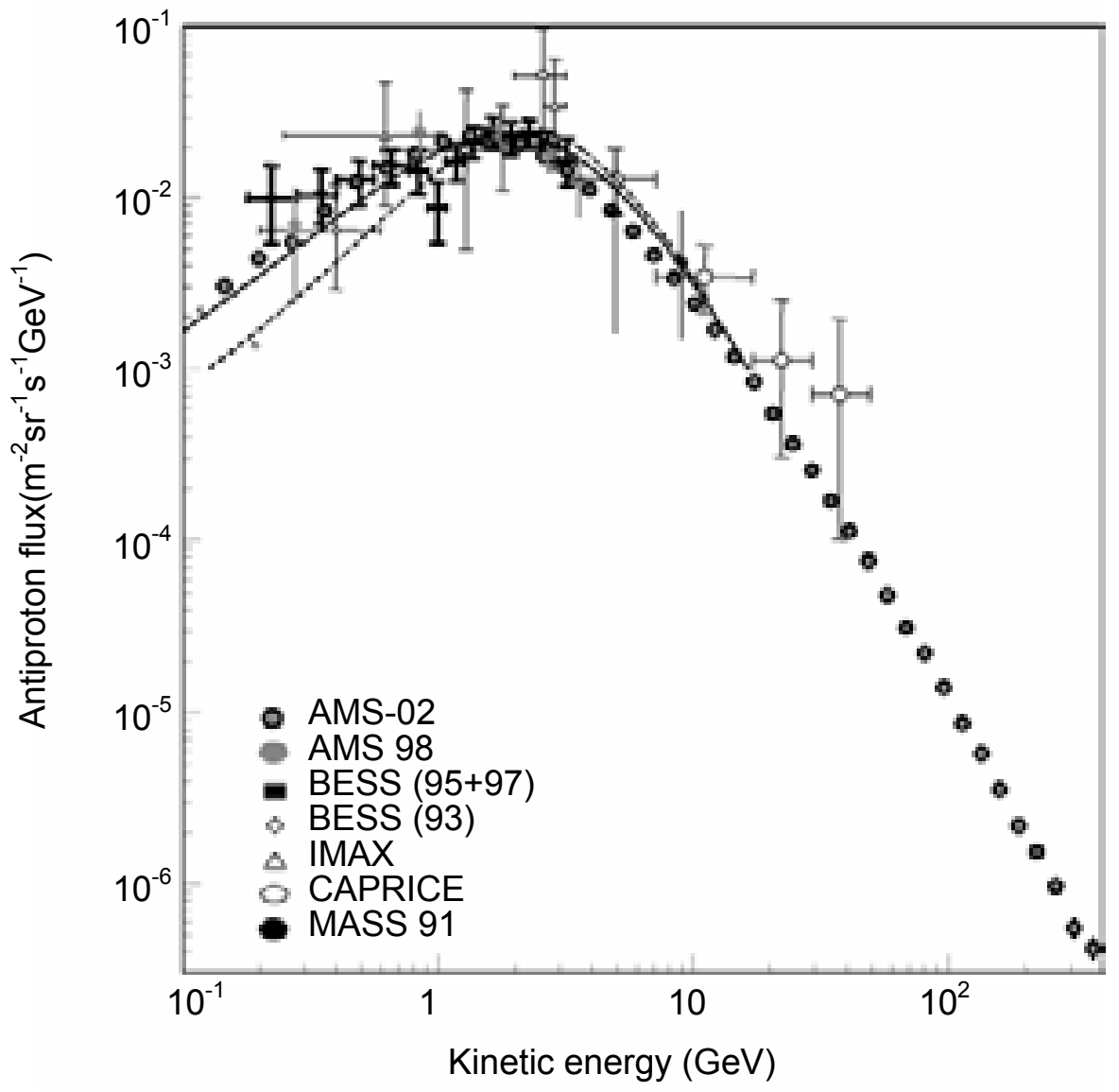

**Fig.22** Antiprotons measured over last 40 years are shown with simulated AMS02 capability to extend the observation window to about 300 GeV with much lower statistic and systematic errors.

### 3.2.3 Electrons and Positrons

There has been a measurement in a balloon-borne experiment (HEAT) with an excess of positrons around 7 GeV over that expected from ordinary sources.[51] However, since there are many other possibilities to create positrons by astrophysical sources the interpretation is not yet conclusive. The accurate measurement of positron spectra could give a clear indication for neutralinos annihilation with rather precise determination of the neutralinos mass.[52]

The interesting energy region is from 2 to about 500 GeV which is accessible to AMS02. AMS02 has an e/p rejection factor of about $10^5$; in three years of data taking will have the statistic error of ~1 % at 50 GeV and ~30 % at 300 GeV with a sensitivity to the exotic fluxes greater than $10^{-7}\ E^2(\mathrm{cm^{-1}\cdot s^{-1}\cdot sr^{-1}\cdot GeV^{-1}})$.

In Fig.23 $e^+/(e^+{+}e^-)$ ratios are compiled for measurements available at present along with the simulated 3 years exposure data of AMS02 from ordinary sources.[53]

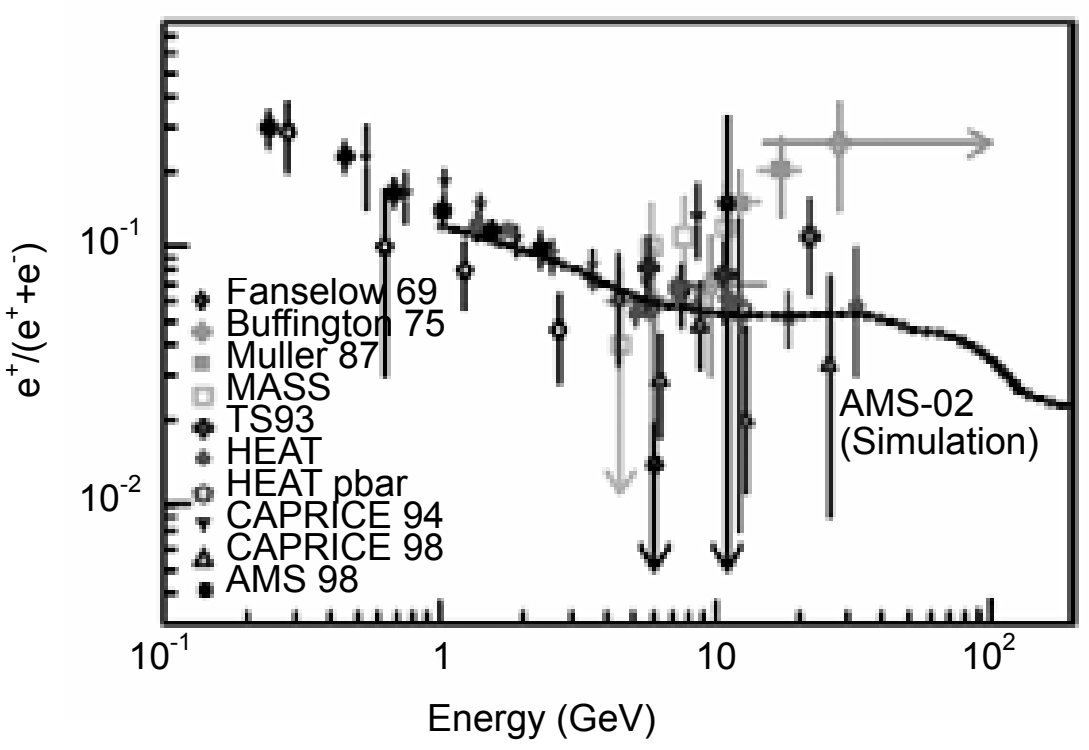

**Fig.23** The summary of 30 years of positron data from balloon (except AMS01). Note the distortion around 7 GeV measured by HEAT experiment.

### 3.2.4 Antideuterons

Antideuteron production from proton-proton collisions is a rare process and it may be less rare in neutralinos annihilation[54] and recently has been pointed out that[55] antideuterons in space could be more promising probe to look for the exotic sources than antiprotons.[56] In particular considering the WIMP pair annihilation in the galactic halo, at energies below about 1 GeV per nucleon the primary antideuteron spectrum would be quite dominant over the secondary one (see Fig.24). The antideuterons measurements open up interesting perspectives for $\chi\chi$ annihilation in space.[57]

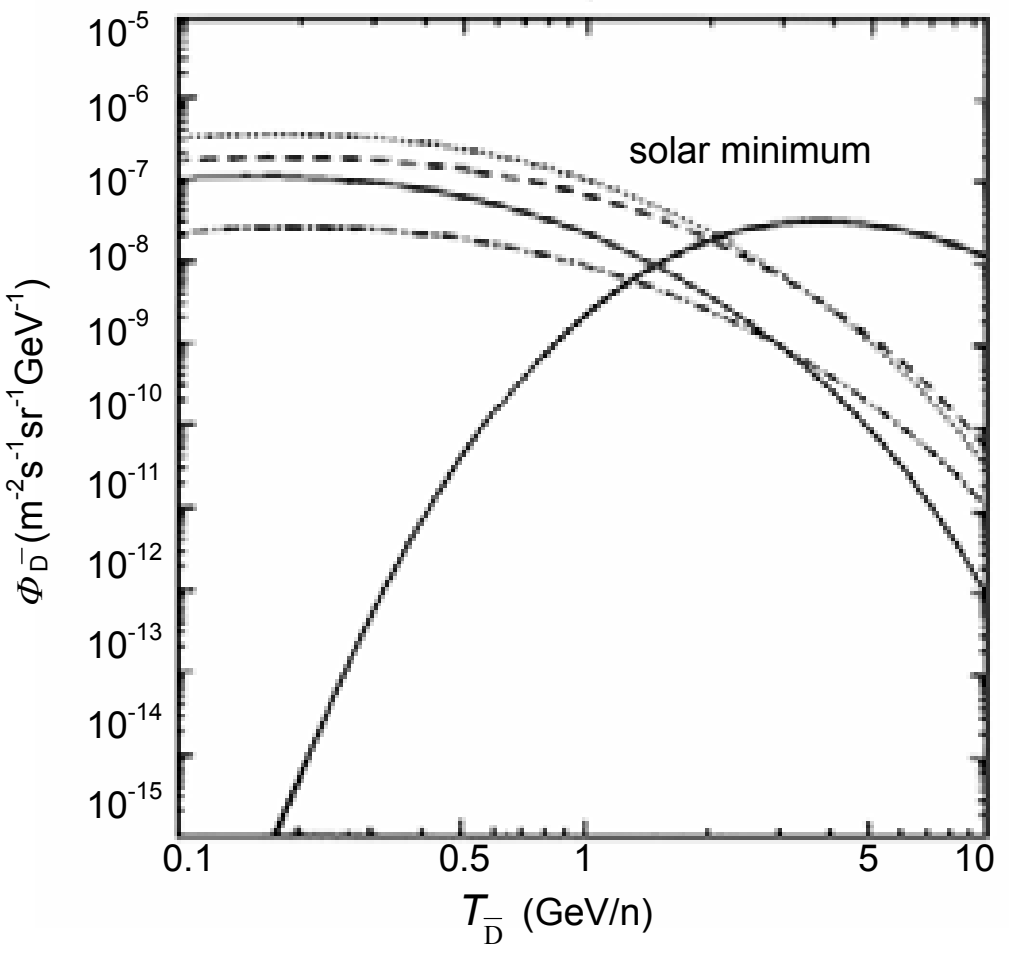

**Fig.24** The Top of the Atmopshere (TOA) antideuterons energy spectra from Ref.[55]. The solid line is for the ordinary production (secondaries) and other four lines at low energies are for four different neutralinos compositions (see [55] for details).

## 3.3 Cosmic ray astrophysics

The study of relative abundances of elements and isotopes will yield to a better understanding of origin, propagation, acceleration and confinement time of cosmic rays in our galaxy.[58] The presence of Earth atmosphere restricts this kind of measurements to be carried out at high altitudes with as low as possible residual atmosphere or better, in space where the effect of Earth's atmosphere is negligible. Nowadays, the experiments aiming to perform elemental and isotopic measurements have suffered limited exposure time, residual atmosphere corrections and relative systematic errors and limited energy ranges.

In galactic cosmic ray propagation models the diffusion coefficient as a function of momentum and the reacceleration are determined by the energy dependence of B/C ratio.[59] The radioactive nuclei data (i.e $^{10}$Be/$^{9}$Be, "radioactive clocks") are used, instead, to derive a range for the height of the cosmic ray halo (3~7 kpc, see [59] for details] as well as to determine the residence time of the galactic cosmic rays in different propagation models.

The AMS02 will be able to measure the particle fluxes with high accuracy to $Z\leqslant25$ in the energy range 0.1 GeV/n$\leqslant$E$\leqslant$1 TeV/n.[60] In Fig.25 the expected B/C ratio for 6 months of AMS02 data is shown together with the present B/C experimental data.

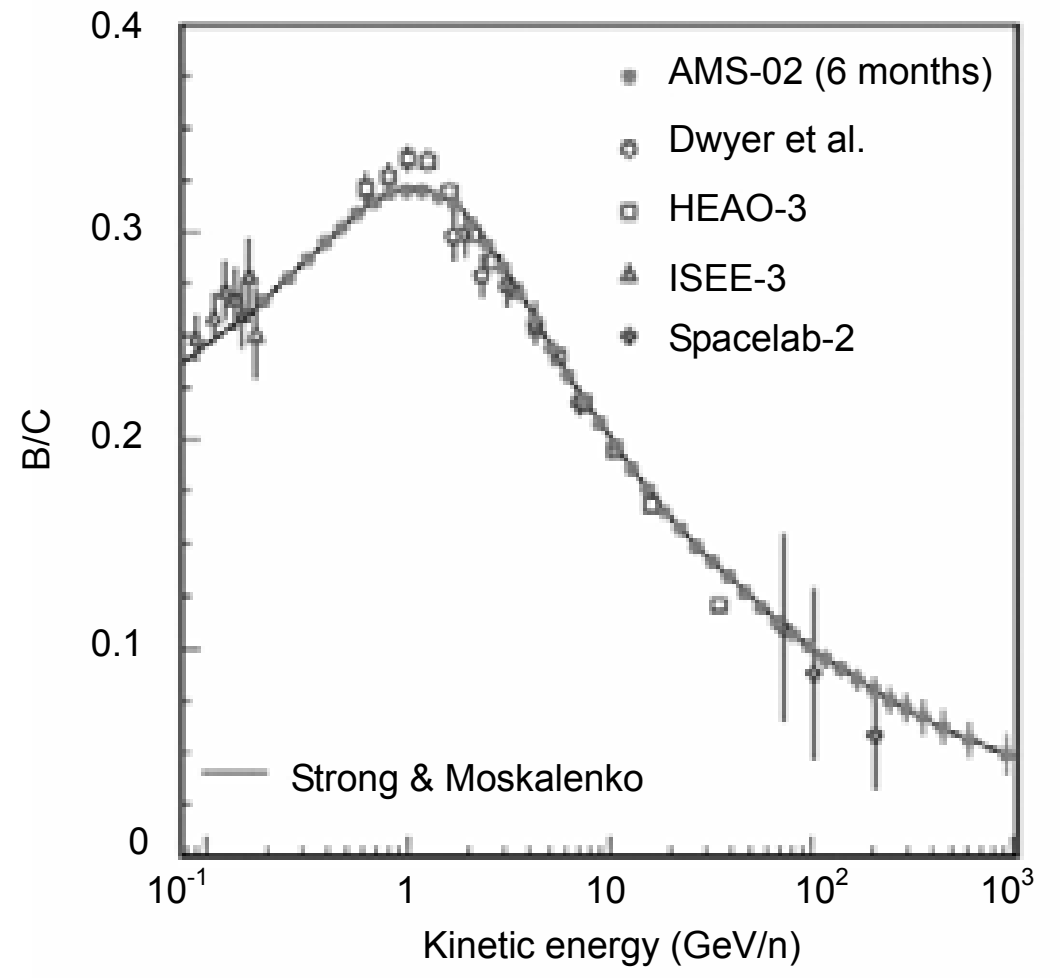

**Fig.25** The B/C ratio with present data and AMS02 after 6 months of data taking (see [60] and references therein) for the models used for simulated AMS02 data .

The AMS02 will separate light isotopes such as $^{2}$H-$^{1}$H and $^{3}$He-$^{4}$He in the energy range of 0.1 GeV/n$\leqslant E \leqslant$10 GeV/n. Accurately measured $^{10}$Be (radioactive with half-life close to the residence time of CR in the galaxy) to $^9$Be (stable) and $^3$He/$^4$He spectral ratio will provide important information to determine the galactic halo size as well as the propagation and diffusion mechanisms of the cosmic rays. In Fig.26 available experimental data are given together with AMS02 expectations for $^3$He/$^4$He and $^{10}$Be/$^9$Be ratios.[60]

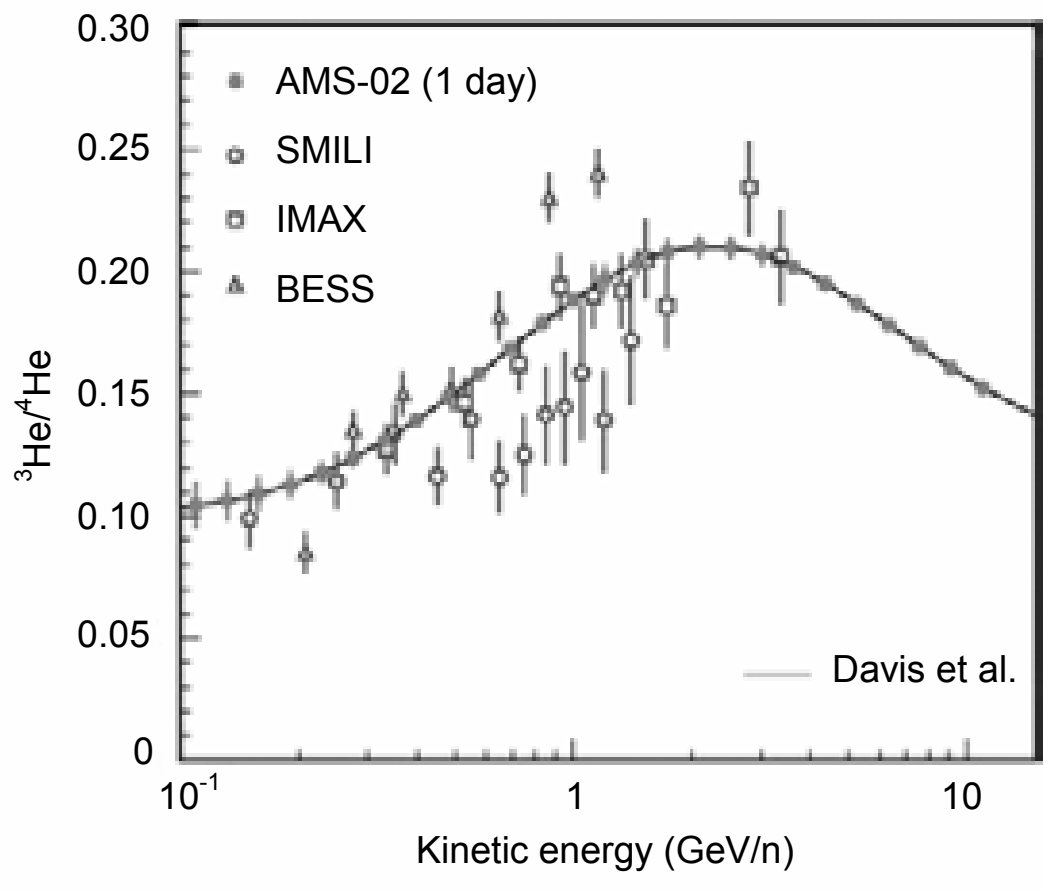

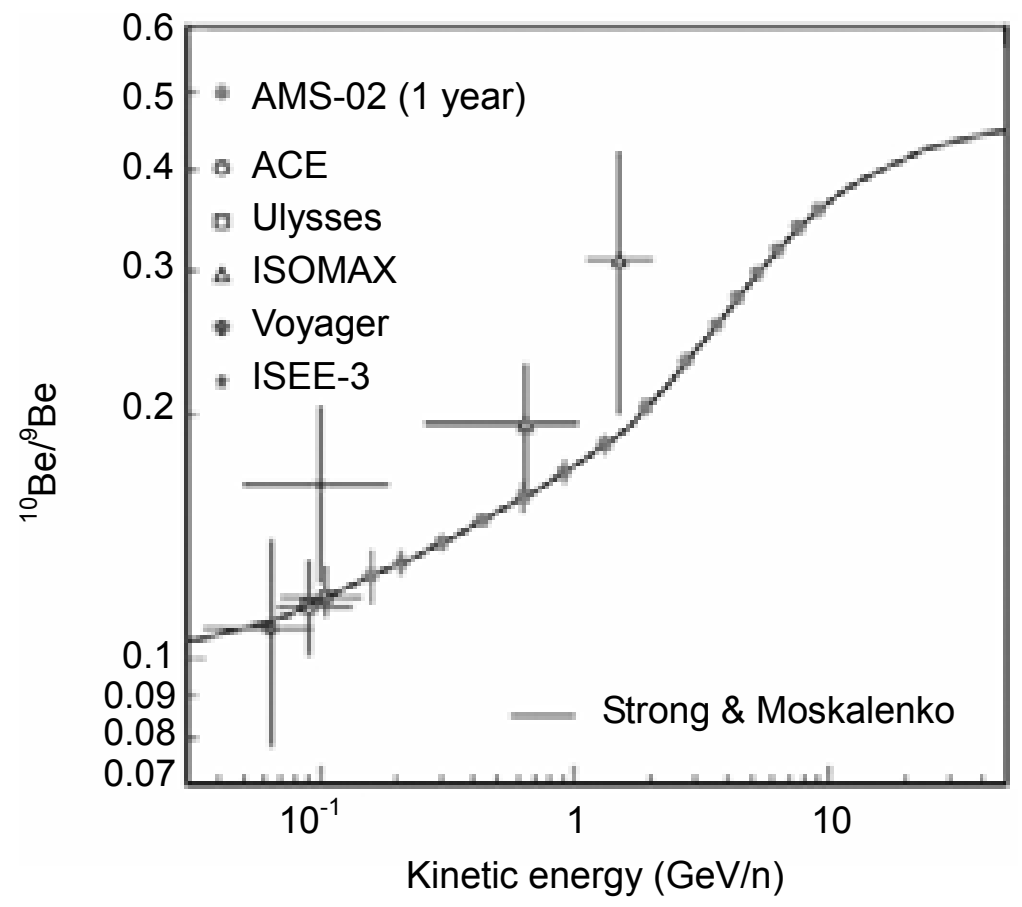

**Fig.26** Available experimental data are given together with the AMS02 expectations for $^3$He/$^4$He (left) and $^{10}$Be/$^9$Be (right) ratios (see [60] and references therein for the models used for simulated AMS02 data).

## 4 Conclusions

The AMS02 is scheduled for installation on the main external truss of the International Space Station in early 2006. Its three years exposure, large acceptance, state-of-art detectors and superconducting magnet will allow accurate measurements of cosmic rays up to the unexplored TeV region.

AMS02 will accurately measure the light element fluxes which are essential for better understanding of cosmic ray origin, propagation, and acceleration mechanisms.

AMS02 will search for cosmological antimatter (antihelium and anticarbon) with unprecedented sensitivity. It is worth underlining that, among those planned for next decade, it will be the unique detector capable of measuring simultaneously four different rare products of $\chi\chi$ annihilation. In addition the AMS02 will open up a window on exotic/other physics such as the study of strangelets.[61, 62]

## Acknowledgements

I wish to thank in particular to Professors Zhiyuan ZHU and Dezhang ZHU, from SINR, giving us the opportunity to prepare the present work and to Professors Cairong ZOU and Qi LI and their colleagues from Southeastern University (Nanjing) for their warm hospitality during our visit in China. I would like to thank also my colleagues R. Battiston and W. J. Burger for fruitful discussions and their careful reading of the draft.